\documentclass[twocolumn,showpacs,preprintnumbers,amsmath,amssymb]{revtex4}

\usepackage{graphicx}
\usepackage{epsfig}
\begin{document}

\title{Dynamics and instability of false vacuum bubbles}

\author{Anthony Aguirre \& Matthew C. Johnson}
\affiliation{
Department of Physics, University of California, Santa Cruz, 
California 95064, USA\\aguirre@scipp.ucsc.edu; mjohnson@physics.ucsc.edu}

\date{\today}

\begin{abstract}
This paper examines the classical dynamics of false vacuum regions
embedded in surrounding regions of true vacuum, in the thin-wall limit.
The dynamics of all generally relativistically allowed solutions -- most
but not all of which have been previously studied -- are derived,
enumerated, and interpreted. We comment on the relation of these solutions
to possible mechanisms whereby inflating regions may be spawned from
non-inflating ones. We then calculate the dynamics of first order
deviations from spherical symmetry, finding that many solutions are
unstable to such aspherical perturbations.  The parameter space in which
the perturbations on bound solutions inevitably become nonlinear is
mapped. This instability has consequences for the Farhi-Guth-Guven
mechanism for baby universe production via quantum tunneling.
\end{abstract}


\maketitle

\section{Introduction}

Nearly two decades ago, a series of papers~\cite{GF87,FMP89,FGG90}
began to investigate the possibility of creating an inflationary
universe ``in a laboratory" -- that is, inside a surrounding region of
much lower vacuum energy.  The spacetime of such a ``bubble universe"
was modeled as a spherically symmetric de Sitter region (the false
vacuum) joined to a surrounding Schwarzschild geometry (the true
vacuum) by an infinitesimally thin domain wall.

These studies found that the inflating (false vacuum) region could not
avoid collapse unless either the null energy condition or cosmic
censorship were violated in the full spacetime~\cite{GF87}, but that
the creation of an enduring inflating region might be possible via
quantum tunneling~\cite{FMP89,FGG90}. In this picture, henceforth
denoted the Farhi-Guth-Guven (FGG) mechanism, a classically
constructable expanding bubble, which would classically re-collapse
(both from the inside and outside perspective~\cite{BGG87}), instead
tunnels to a new solution in which the inflating interior expands
forever, while an outside observer sees a black hole~\footnote{We
will include the means by which the constructable expanding bubble is
produced, be it in the lab or by some quantum or thermal fluctuation,
in the definition of the Farhi-Guth-Guven mechanism}. The probability
of this occurring can be calculated using the techniques of
semi-classical quantum gravity, and is extraordinarily small.

Though nearly-miraculous, this process has garnered new interest
recently, primarily because of evidence that our universe may have a
fundamental positive cosmological constant.  If so, it will
asymptotically approach everlasting equilibrium as de Sitter
spacetime.  Given eternity, even the most unlikely process -- such as
the creation of inflating bubble universes -- will eventually occur.
Taking this one step further, our observable universe could in fact be
such a bubble universe, which arose from equilibrium de Sitter space
and is currently returning to it.  This would realize the old idea of
Boltzmann that the universe is fundamentally in equilibrium, but that
extremely rare downward fluctuations in entropy periodically occur and
allow the transitory existence of non-equilibrium regions that see
entropy steadily increasing.

The classic problem with this idea was pointed out in its new context
by Dyson, Susskind, \& Kleban~\cite{DKS02}: if our observable universe
resulted from a downward entropy fluctuation that evolved with
increasing entropy to the present time, then it is vastly more likely
for this to have occurred by a fluctuation to our observable universe
ten minutes ago (replete with incoming photons and memories in our
brains to convince us that it is older) than by a fluctuation all the
way to the much lower entropy corresponding to
inflation~\footnote{This, in fact, connects to a more general question
of whether inflation can really be said to have ``general'' initial
conditions, given that it must be low entropy; see,
e.g.,~\cite{penrose,Hollands:2002yb,Carroll:2004pn}.}. Albrecht and
Sorbo~\cite{AS04}, however, argue that inflation might avoid this
problem by turning a tiny region of low entropy {\em density} into a
very large one. Their calculation shows that the FGG mechanism
requires a much smaller entropy fluctuation than directly creating the
large post-reheating region that will result from it, thus resolving
the paradox.

There may, however, be reasons to doubt that the creation of small
regions of false vacuum and subsequent tunneling are plausible
events. First, there is no regular instanton describing either the
nucleation of small regions of false vacuum or the gravitational
tunneling event; there {\em is} such an instanton describing tunneling
to false-vacuum, but over a huge region, larger than the true-vacuum
horizon (see, e.g.,~\cite{LW87,G94}). Second,
Banks~\cite{Banks:2002nm} has pointed out that it is hard to
understand the tunneling process holographically -- it appears that
the inflating region inside the bubble has many more states than the
black hole it is ``contained in" (see
also~\cite{Bousso:2004tv}). Third, Dutta \& Vachaspati~\cite{DV05}
have recently (and previously in~\cite{VT99}) argued that general
causality considerations preclude the formation of a small
false-vacuum region inside a large true-vacuum one.

Both in terms of the initial conditions for inflation, and also the
more general issue of what processes can lead to transitions between
vacuum states (as is important in understanding the string theory
``landscape''), it is crucial to understand bubble universes, whether
they can form, and with what probabilities.  To this question the
present paper makes the following contributions: first, we organize
and interpret all of the thin-walled, spherically symmetric one bubble
spacetimes.  We then show that expanding false vacuum bubbles are
unstable to non-spherical perturbations; {\em if} these bubbles start
at a sufficiently small initial radius, then they inevitably become
nonlinearly aspherical before tunneling might occur in the FGG
mechanism. It is unclear in this case whether tunneling to an
inflationary universe inside the bubble can occur at all, or with what
probability.

The plan of the paper is as follows. In Sec.~\ref{intro}, the allowed
solutions to the junction conditions are enumerated, putting the
previous work into context. We provide a concise reference for all of
the possible spacetimes with one false-vacuum bubble and arbitrary
positive cosmological constant, then discuss the existing, and some new,
interpretations of these solutions. In Sec.~\ref{perts} we derive the
first order perturbation equations, and demonstrate the existence of
an instability in bound solutions. In Sec.~\ref{sec-results} we
integrate the equations to investigate the parameter space for which
non-linear perturbations are unavoidable, and we conclude in
Sec.~\ref{conclusions}.

\section{Junction Conditions  \label{intro}}
The dynamics of inflating regions has been discussed by a number of
authors~\cite{BGG87,BKT87,S86,APS89}\footnote{We follow the methods and
notations of these works, though they are not the only on the
subject.} using the junction condition formalism. These works study a
spherically symmetric region of false vacuum (high energy density) in
a surrounding region of true vacuum (lower energy density). The wall
separating the regions is assumed to be very thin compared to the
radius of the region of false vacuum. One can obtain the dynamics of
the wall by requiring metric continuity across the wall and then
solving Einstein's equations. Under spherical symmetry, the dynamics
of the problem then reduce to those of the bubble wall's radius.  This
radius is a gauge invariant quantity because it simply quantifies the
curvature of the bubble wall worldsheet, and any observer can measure
it by comparing the normal to the wall at two nearby points.

\subsection{Interior and Exterior Spacetimes}

Let $\Lambda_{+}$ be the cosmological constant in the true vacuum
region.  Then if $\Lambda_{+} > 0$, the region is Schwarzschild--de
Sitter spacetime with metric
\begin{equation}\label{gsds}
ds_{+}^2=-a_{\rm sds}dt^2 + a_{\rm sds}^{-1}dr^2 + r^2 d\Omega^2,
\end{equation}
\begin{equation}\label{defa}
a_{\rm sds}=1-\frac{2M}{r} - \frac{\Lambda_{+}}{3} r^2
\end{equation}
in the static foliation. Fixing $\Lambda_{+}$, there are then three
qualitatively different casual structures characterized by the value
of $M$ (see \cite{LR77}), due to the nature of the three roots of
$a_{\rm sds}(r).$

\begin{subequations}
For $3M<\Lambda_{+}^{-1/2}$, there are three distinct real roots of form:
\begin{equation}\label{rbh}
r_n=2 (\Lambda_{+})^{-1/2} \cos{\left(\frac{\theta}{3} +\frac{2\pi n}{3}\right)},
\end{equation}
where 
\begin{equation}
\cos \theta = -3M (\Lambda_{+})^{1/2},
\end{equation}
\end{subequations}
and  $\pi<\theta<3\pi/2$.
We can label them as 
$$r_{BH}\equiv r_0,\ \ r_{\rm neg}=r_1,\ \ r_c=r_2,$$ And the range of
$\theta$ means that they lie in the ranges
$r_{\rm neg}<0<2M<r_{BH}<3M<r_{c}$.

\begin{figure}
\centering
\includegraphics[width=8.6cm]{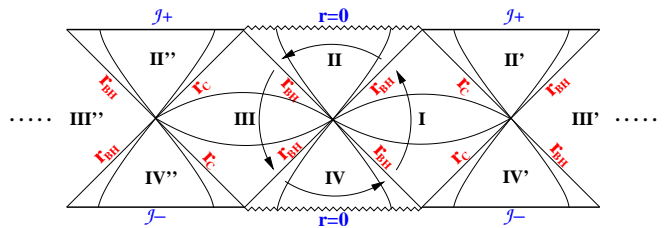}
\caption{Conformal diagram of the Schwarzschild de
Sitter geometry for $3M<\Lambda_{+}$. \label{sdsconf}}
\end{figure}

The two positive roots correspond to the black hole and cosmological
horizons. The conformal diagram for this spacetime is shown in
Fig.~\ref{sdsconf}. (See \cite{CP04} for a demonstration of the
explicit form of the metric in global coordinates). Surfaces of
constant coordinate time $t$ are drawn, with the circulating arrows
denoting the direction of increasing $t$. We will consider region I to
be the ``causal patch'' of a hypothetical observer (i.e., the region
lying in both the causal past and causal future of the observer's
world line) in what follows.

\begin{figure}
\centering
\includegraphics[width=6cm]{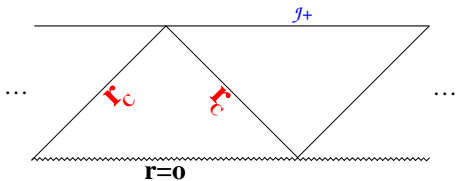}
\caption{Conformal diagram for the Schwarzschild de
Sitter geometry when $3M=\Lambda_{+}$. \label{sdsconf2}}
\end{figure}

For $3M=\Lambda_{+}^{-1/2}$, there are also three real roots: a double
positive root $r_h$ and a negative $r_{\rm neg}$, given by:
\begin{equation}
r_{h} = \Lambda_{+}^{-1/2}, \ \ r_{\rm neg}=-2 \Lambda_{+}^{-1/2}.
\end{equation} 
This mass is known as the Nariai mass, and in this spacetime there is
only one horizon at the positive root. The conformal diagram for this
spacetime is shown in Fig.~\ref{sdsconf2} \cite{LR77}. There
is also a time-reverse solution, starting at past null infinity and
ending at $r=0$. For $3M>\Lambda_{+}^{-1/2}$, there is one real negative
root, and therefore no horizons in the spacetime.  The conformal
diagram for this case is like figure \ref{sdsconf2}, but with the
horizon lines excised.

\begin{figure}
\centering
\includegraphics[width=3.5cm]{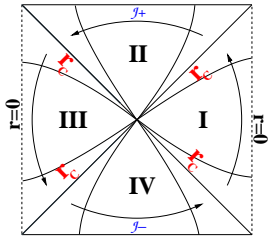}
\caption{Conformal diagram for the de Sitter geometry.\label{dsconf}}
\end{figure}

Inside the false-vacuum region the spacetime is de Sitter space, with
metric
\begin{equation}
 ds_{-}^2 = -a_{\rm ds} dt^2 + a_{\rm ds}^{-1} dr^2 +r^2 d\Omega^2,
\end{equation}
\begin{equation}
a_{\rm ds}=1- \frac{\Lambda_{-}}{3}r^2
\end{equation} 
in the static foliation. Fig.~\ref{dsconf} shows the conformal diagram
for the de Sitter region. Again, surfaces of constant coordinate time
$t$ are shown, with the arrows denoting the direction of increasing
$t$. We consider region III to be the causal patch in which our
hypothetical false vacuum observer resides.

\subsection{Equation of Motion}\label{junctioneom}
The bubble wall worldsheet has metric:
\begin{equation}\label{wallmetric}
ds^2= - d\tau^2 + r(\tau)^2 d\Omega^2,
\end{equation}
where $\tau$ is the proper time in the frame of the wall, and
($\theta, \phi$) are the usual angular variables.

The coordinates in the full $4D$ spacetime are chosen to be Gaussian
normal coordinates constructed in the neighborhood of the bubble wall
worldsheet. Three of the coordinates are $(\tau,\theta,\phi)$ on the
worldsheet, and the fourth, $\eta$, is defined as the proper distance
along a geodesic normal to the bubble worldsheet, with $\eta$
increasing in the direction of SdS (true vacuum).

The transformation from the static coordinate systems of SdS and dS to
the Gaussian normal system can be constructed in closed form using the
methods of \cite{FGG90}, and the full metric takes the form:
\begin{equation}
\label{eq-4dmetric}
ds^2 = g_{\tau\tau}(\tau,\eta) d\tau^2 + d\eta^2 + R(\tau,\eta)^2
d\Omega^2,
\end{equation}
where $\eta=0$ defines the wall and therefore
$g_{\tau\tau}(\tau,0)=-1$ and $R(\tau,0)=r(\tau)$.

The energy momentum tensor on the wall is:
\begin{equation}
\label{eq-emtens}
T^{\mu\nu}_{\rm wall} = -\sigma \gamma^{\mu\nu} \delta(\eta)
\end{equation}
where $\gamma^{\mu\nu}$ is the metric on the worldsheet of the wall
for $\mu=\nu=\tau,\theta,\phi$ and zero otherwise, and
$\sigma$ is the energy density of the wall.
 
Using the metric~\ref{eq-4dmetric} and the energy-momentum
tensor~\ref{eq-emtens} together with the contributions from the dS
interior and SdS exterior in Einstein's equations yields an equation
of motion for the bubble wall of~\cite{FGG90,BGG87} :
\begin{equation}
 \label{preisrael}
K^{i}_{j}(\eta_{+}) - K^{i}_{j}(\eta_{-}) = - 4 \pi \sigma r
\delta^{i}_{j},
\end{equation}
where $K^{i}_{j}(\eta_{\pm})$ is the extrinsic curvature tensor in SdS
and dS respectively. In the Gaussian normal coordinates, this takes
the form:
\begin{equation}
K_{ij}=\frac{1}{2} \frac{d}{d\eta}g_{ij}
\end{equation}

Evaluating this in metric~\ref{eq-4dmetric}, the $\theta\theta$ and
$\phi\phi$ components of Eq.~\ref{preisrael} reduce to:
\begin{equation}
 \label{israel}
\beta_{\rm ds} - \beta_{\rm sds} = 4 \pi \sigma r,
\end{equation}
with the definitions
\begin{equation}
\beta_{\rm ds} \equiv - a_{\rm ds} \frac{dt}{d\tau},\ \ \beta_{\rm sds} \equiv
a_{\rm sds} \frac{dt}{d\tau}.
\end{equation}
Here, $a$ is the metric coefficient in dS or SdS. The sign of $\beta$
is fixed by the trajectory because $dt/d\tau$ could potentially be
positive or negative (motion can be with or against the direction of
increasing coordinate time indicated in Fig.\ref{sdsconf} and
\ref{dsconf}).

A set of dimensionless coordinates can be defined, in which
Eq.~\ref{israel} can be written as the equation of motion of a
particle of unit mass in a one dimensional
potential~\cite{APS89}. Let:

\begin{equation} \label{ztor}
z=\left(\frac{L^2}{2M}\right)^{\frac{1}{3}}r,\ \ T = \frac{L^2}{2k} \tau,
\end{equation}
where M is the mass appearing in the SdS metric coefficient, and 
\begin{equation}
k=4\pi\sigma,
\end{equation}
\begin{equation}\label{Lsq}
L^2=\frac{1}{3}\left[\left| \left(\Lambda_{-} + \Lambda_{+} + 3k^2 \right)^2 - 4\Lambda_{+}\Lambda_{-}\right| \right]^{\frac{1}{2}}.
\end{equation}
With these definitions, Eq.~\ref{israel} becomes 
\begin{equation}\label{juncteom}
\left[\frac{dz}{d T }\right]^2=Q-V(z),
\end{equation}
where the potential $V(z)$ and energy $Q$ are 
\begin{equation}\label{potential}
V(z)=-\left[z^2+\frac{2Y}{z}+\frac{1}{z^4} \right],
\end{equation}
with
\begin{equation} \label{y}
Y=\frac{1}{3}\frac{\Lambda_{+}-\Lambda_{-}+3k^2}{L^2},
\end{equation}
and
\begin{equation} \label{Qtom}
Q=-\frac{4k^2}{\left(2M\right)^{\frac{2}{3}}L^{\frac{8}{3}}}.
\end{equation}
Note that a small negative $Q$ corresponds to a large mass, so that
even between $-1<Q<0$ the mass will vary by many orders of magnitude.

\begin{figure}
\centering
\includegraphics[width=8cm]{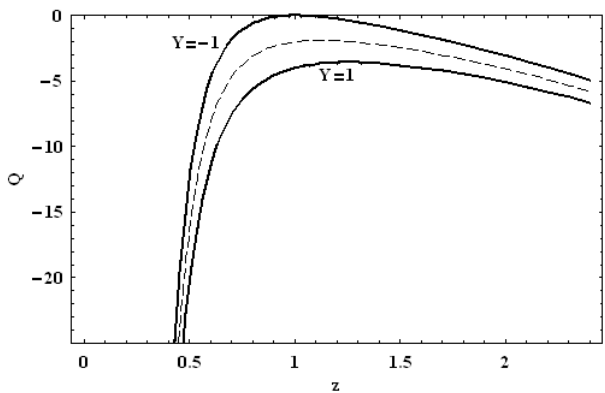}
\caption{The potential for various Y. \label{potvsY}}
\end{figure}

The parameter space allowed by the junction conditions is
characterized by the value of the cosmological constant inside and
outside the wall. For $0 \leq \Lambda_{+} < \Lambda_{-}$, we have $-1
\leq Y \leq 1$. The maximum $V_{\rm max}$ of the potential $V(z)$ then
satisfies $-2^{5/3}-2^{-4/3} \leq V_{max} \leq 0$. The potential
curves over the entire range of $Y$ are shown in Fig.\ref{potvsY}.

The interior and exterior cosmological constants can be expressed in
terms of $k^2$ as $\Lambda_{+}=Ak^2$ and $\Lambda_{-}=Bk^2$. With
these choices, the dynamics of the bubble wall are entirely determined
by $A$, $B$, and $Q$.

Let us now discuss some realistic values for the parameters in this
theory. The interior cosmological constant ($\Lambda_{-} =
M_{I}^{4}/M_{pl}^{4}$) and the bubble wall surface energy density ($k
= 4 \pi M_{I}^{3}/M_{pl}^{3}$) will be set by the scale of inflation
($M_{I}$). The exterior cosmological constant ($\Lambda_{+} =
M_{\Lambda}^{4}/M_{pl}^{4}$) will be set by a scale
$M_{\Lambda}$. These yield
\begin{equation}\label{scaleAB}
A = \frac{M_{\Lambda}^{4} M_{\rm pl}^{2}}{(4 \pi)^2 M_{I}^{6}},\ \ 
B = \frac{ M_{\rm pl}^{2}}{(4 \pi)^2  M_{I}^{2}}.
\end{equation}

We will consider three representative energy scales, covering the
interesting range of energy scales for inflation. For weak scale
inflation ($100$ GeV), $k \simeq 4 \pi \times 10^{-51}$, $A \simeq 0$,
and $B \simeq 10^{32}$. For an inflation scale near the GUT scale
($10^{14}$ GeV $\gg M_{\Lambda}$), we have $k \simeq 4 \pi \times
10^{-15}$, $A \simeq 0$, and $B \simeq 10^7$. Near-Planck scale
inflation ($10^{17}$ GeV) yields $k \simeq 4 \pi \times 10^{-6}$, $A
\simeq 0$, and $B \simeq 63$. The most massive bound solution (that
which just reaches the top of the potential) is given by converting
from $Q$ to $M$ using Eq.~\ref{Qtom}. This maximal mass is very
different in each case, ranging from an ant-mass of $M_{\rm max}
\simeq 10^3 M_{\rm pl}\simeq 10^{-2}$ grams for $M_{I}=10^{17}$ GeV to
an Earth-mass of $M_{\rm max} \simeq 10^{33}M_{\rm pl} \simeq 10^{28}$
grams, for $M_{I}=100$ GeV.

\subsection{Allowed Solutions}\label{sols}
A bubble wall trajectory is characterized by $Q=$const., and there are 
three general types:

\begin{itemize}
\item Bound solutions with $Q < V_{\rm max}$. These solutions start at
  $z=0$, bounce off the potential wall and return to $z=0$.
\item Unbound solutions with $Q < V_{\rm max}$. These solutions
start at $z=\infty$, bounce off the potential wall and return to
$z=\infty$.
\item Monotonic solutions with $Q > V_{\rm max}$. These solutions
start out at $z=0$ and go to $z=\infty$, or execute the time-reversed motion. 
\end{itemize}

\begin{figure}
\centering
\includegraphics[width=8.6cm]{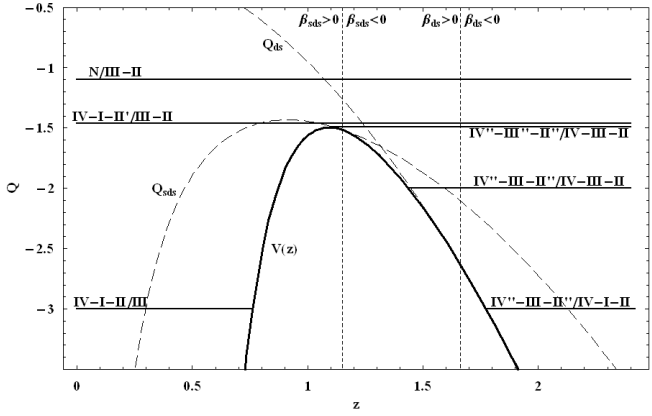}
\caption{Potential for $A=9$ and $B=15$. The two dashed lines labeled
$Q_{\rm sds}$ and $Q_s$ represent the exterior and interior horizon
crossings respectively. The vertical dotted lines denote the regions
in which $\beta_{\rm sds}$ and $ \beta_{\rm ds}$ are positive and
negative. Various trajectories are noted. \label{apspotential}}
\end{figure}

\begin{figure}
\centering
\includegraphics[width=8.6cm]{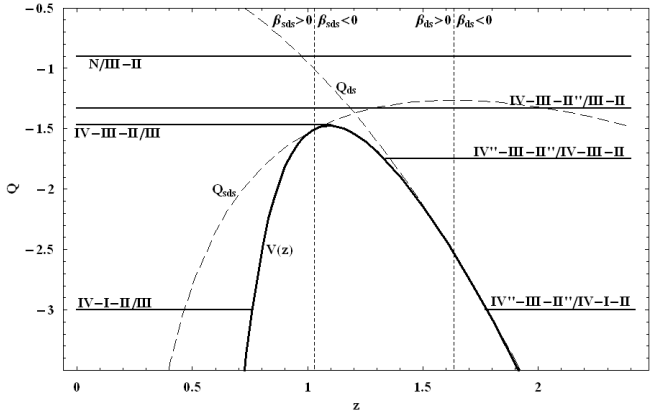}
\caption{Potential for $A=1$ and $B=6$. For this choice of parameters,
the sign change in $\beta_{\rm sds}$ occurs to the left of the maximum
in the potential. Various trajectories are noted. \label{gfpotential}}
\end{figure}

\begin{figure}
\centering
\includegraphics[width=8.6cm]{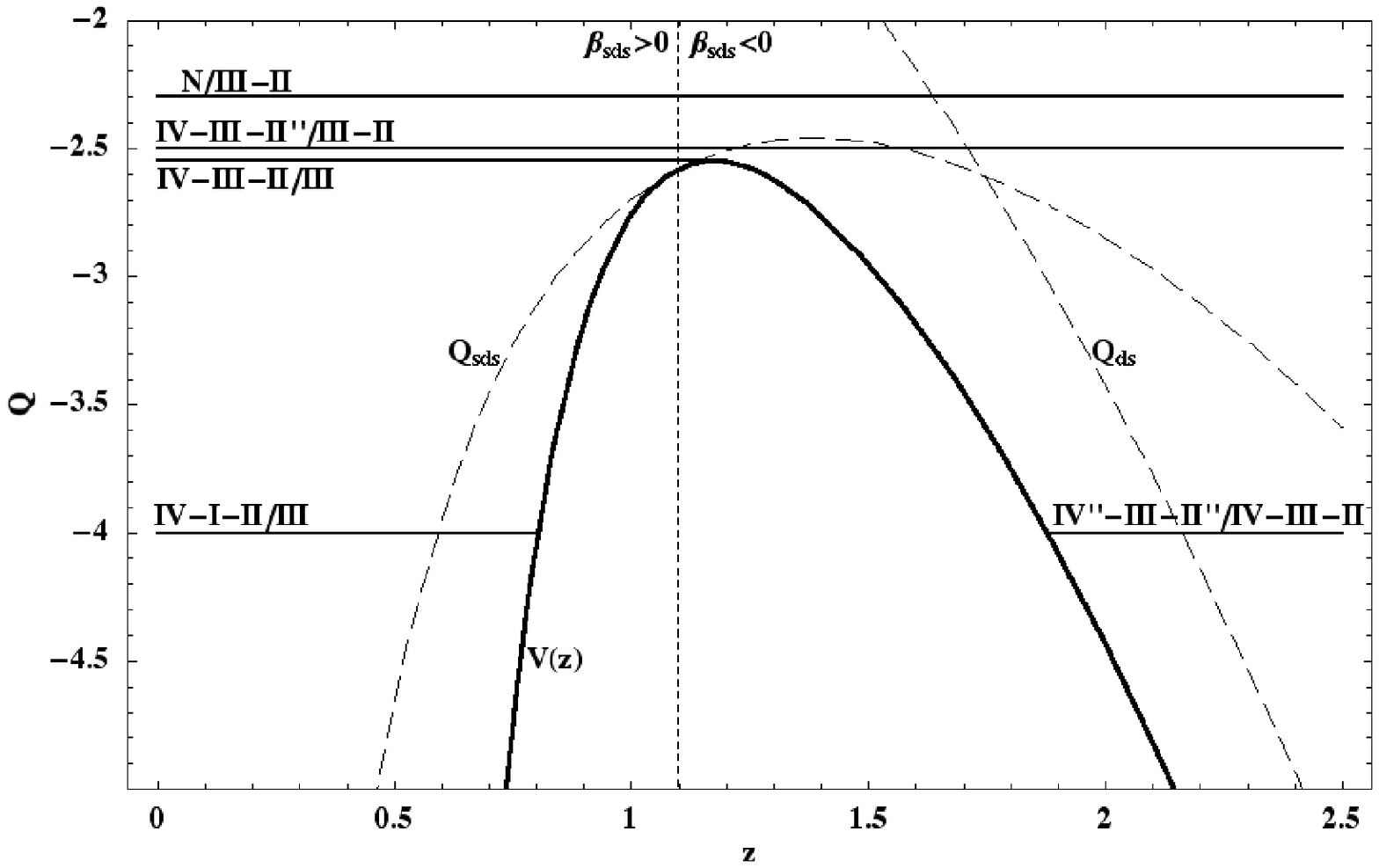}
\caption{Potential for $A=1$ and $B=2$. For this choice of parameters,
the sign change in $\beta_{\rm sds}$ occurs to the left of the maximum
in the potential and there is no $\beta_{\rm ds}$ sign change. Various trajectories are noted. \label{A1B2}}
\end{figure}

\begin{figure}
\centering
\includegraphics[width=8.6cm]{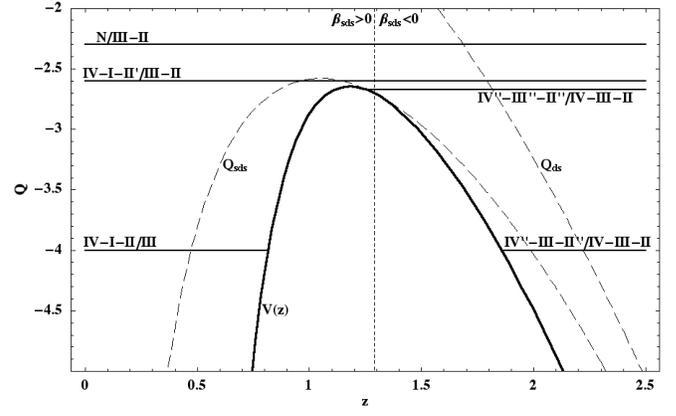}
\caption{Potential for $A=2.9$ and $B=3$. For this choice of parameters,
the sign change in $\beta_{\rm sds}$ occurs to the right of the maximum
in the potential and there is no $\beta_{\rm ds}$ sign change. Various trajectories are noted. \label{A2_9B3}}
\end{figure}

The qualitative features of a generic potential can be shown by
considering the four illustrative (but unrealistic; see above) cases of
($A=9$, $B=15$) shown in Fig.~\ref{apspotential}, ($A=1$, $B=6$)
shown in Fig.~\ref{gfpotential}, ($A=1$, $B=2$) shown in Fig.~\ref{A1B2}, and ($A=2.9$, $B=3$) shown in Fig.~\ref{A2_9B3}. The important features are:

\begin{itemize}
\item 
As one follows a line of constant $Q$ in Fig.~\ref{apspotential}, \ref{gfpotential}, \ref{A1B2}, and \ref{A2_9B3} every intersection with the dashed line $Q_{\rm sds}$ (which is obtained by solving $a_{\rm sds} = 0$ for $Q$) 
represents a horizon crossing in the SdS spacetime (this could either
represent the black hole or cosmological horizons). 
\item Intersections with the dashed line $Q_{\rm ds}$ (which is obtained by solving $a_{\rm ds} = 0$ for $Q$) as one moves along a line of constant $Q$ represent the crossing of the interior dS horizon.
\item The vertical line on the right denotes the position at which
$\beta_{\rm ds}$ changes sign. $\beta_{\rm ds} > 0 $ if $t_{\rm ds}$ is
decreasing along the bubble wall trajectory and is negative if
$t_{\rm ds}$ is increasing. For there to be a $\beta_{\rm ds}$ sign change, $Y$ in Eq.~\ref{y} must be in the range $-1 \leq Y < 0$ \cite{APS89}, which yields the condition that $B > A+3$ for a sign change to occur.
\item The vertical dotted line on the left denotes the radius at which
$\beta_{\rm sds}$ changes sign. Recall that $\beta_{\rm sds} > 0$ if
$t_{\rm sds}$ is increasing along the bubble wall trajectory, and
$\beta_{\rm sds} < 0 $ if it is decreasing. For the parameters in
Fig.~\ref{apspotential} and \ref{A2_9B3}, or (one can show) whenever $B<3(A-1)$, the
sign change occurs to the right of the maximum of $V$.  When
$B>3(A-1)$ it occurs to the left~\footnote{Ref. \cite{APS89}
considered only the case where the sign change occurred to the left of
the maximum.}, as shown in Fig.~\ref{gfpotential} (note that this is the
interesting case where $\Lambda_{+} \ll \Lambda_{-}$) and \ref{A1B2}.
\end{itemize}

\begin{figure*}
\includegraphics[width=16.4cm]{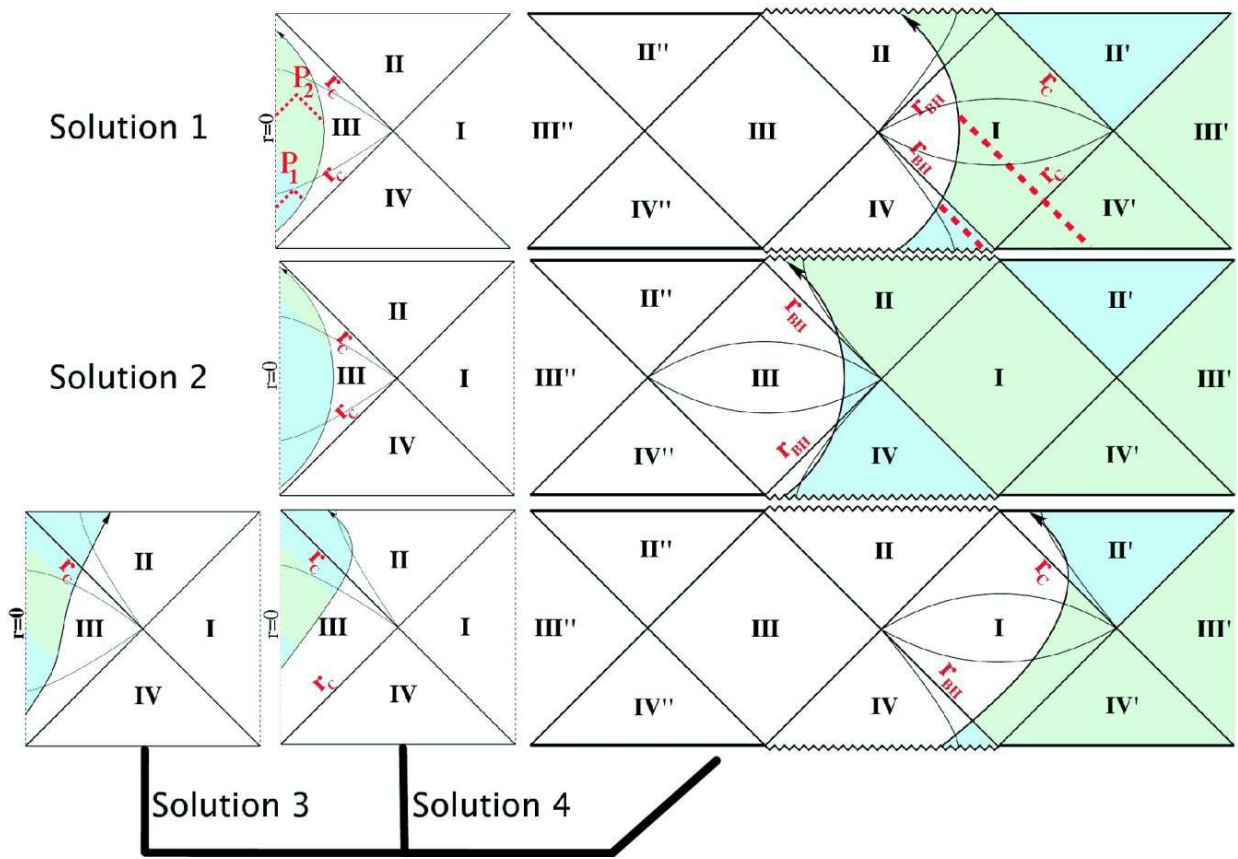}
\caption{Conformal diagrams for the allowed one-bubble spacetimes. The dS and SdS diagrams are matched across the bubble wall (line with
arrow), and the physical regions shaded. Solutions 3 and 4 share the same SdS diagram. Regions which
do not contain anti-trapped surfaces are shaded green (light), regions
which do are shaded blue (dark).
\label{confdiags1}}
\end{figure*}

\begin{figure*}
\includegraphics[width=17.4cm]{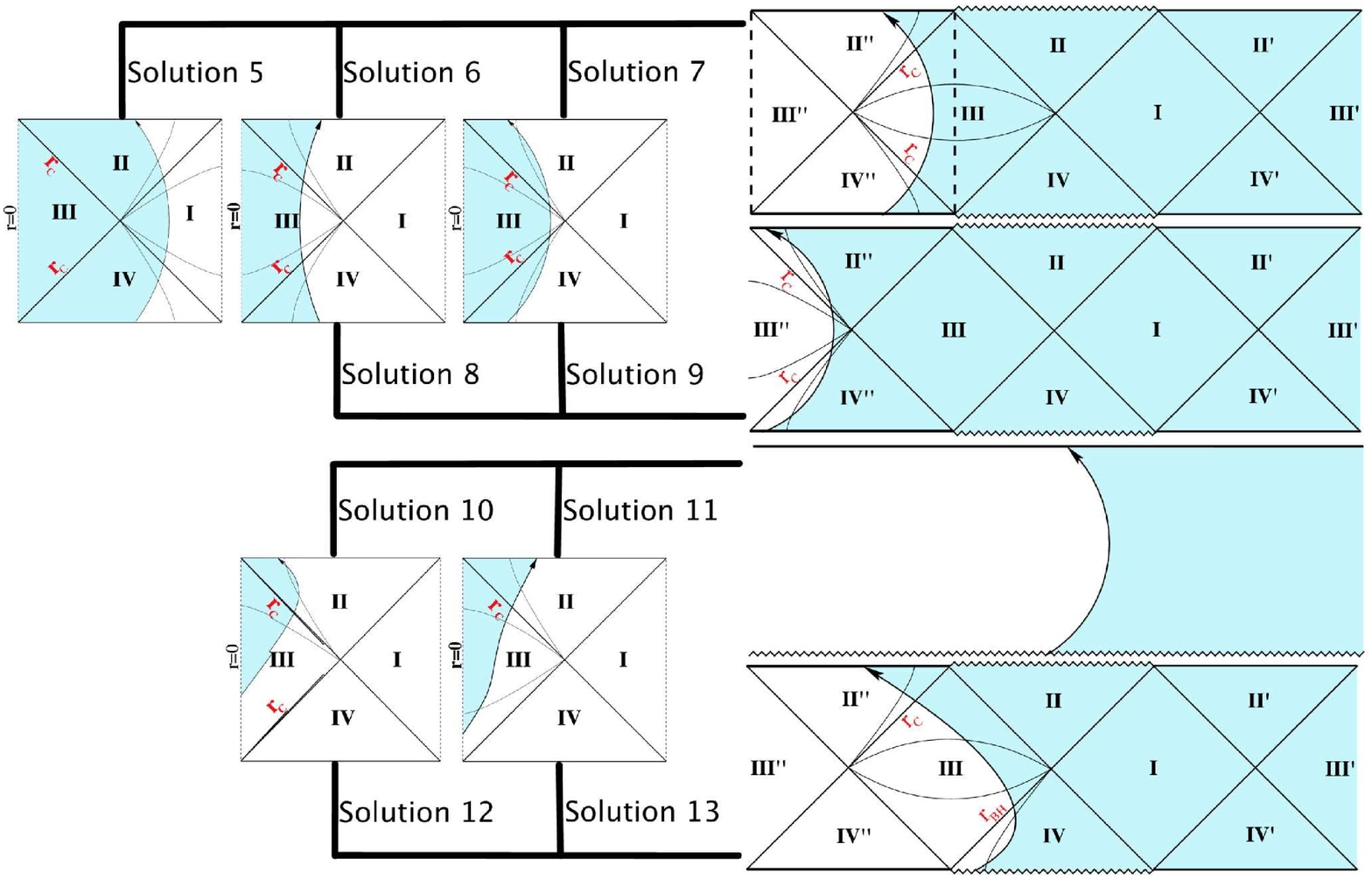}
\caption{Conformal diagrams for the allowed one-bubble spacetimes. The dS and SdS diagrams are matched across the bubble wall (line with arrow), and the physical regions shaded. For solutions with qualitatively similar SdS diagrams, the various options for the dS interior are listed.
\label{confdiags2}}
\end{figure*}

The potential diagram contains all of the information needed to
determine the conformal structure of the allowed one-bubble
spacetimes. A complete set of the qualitatively different trajectories
for arbitrary interior and exterior positive definite cosmological
constants (with $\Lambda_{+} < \Lambda_{-}$) and $M\ge 0$ are denoted
in Fig.~\ref{apspotential}, \ref{gfpotential}, \ref{A1B2}, and \ref{A2_9B3}. Figures~\ref{confdiags1} and~\ref{confdiags2} display the conformal structure of these solutions \footnote{Many, but not all, of these solutions have appeared previously in the
literature~\cite{GF87,APS89,BKT87}}. The conformal diagrams in each
row are matched along the bubble wall (solid line with an arrow) and
the physical regions are shaded. For solutions with qualitatively similar SdS diagrams, the various options for the dS interior are listed. The naming scheme in
Fig.~\ref{apspotential}, \ref{gfpotential}, \ref{A1B2}, and \ref{A2_9B3} is chosen to reflect the structure of the conformal diagrams. The first numbers are the regions
of the SdS conformal diagram that the bubble wall passes through. The
numbers following the backslash are the regions of the dS conformal
diagram that the bubble wall passes through.

For example, consider the IV-I-II/III solution (solution 1 of Fig.~\ref{confdiags1}). These start
at at $r=z=0$, corresponding to the singularity in region IV of the
SdS conformal diagram and the $r=0$ surface in region III of the dS
conformal diagram. $\beta_{\rm sds}$ and $\beta_{\rm ds}$ are both
greater than zero over the entire trajectory. Therefore, the wall on
the SdS side must follow a path of increasing $t_{\rm sds}$, pushing
it into region I (crossing the black hole horizon). The wall on the dS
half must follow a path of decreasing coordinate time, and thus
remain in region III. The wall then reaches a turning point, falls
through the black hole event horizon, and ends up back at $r=z=0$ (the
singularity in region II of the SdS diagram, and the $r=0$ surface in
region III of the dS diagram). The construction of the other diagrams
in Fig.~\ref{confdiags1} and Fig.~\ref{confdiags2} proceeds similarly.

Based on the inward pressure gradient, the IV-I-II/III solution is
what one might expect the junction conditions to yield: a sphere of
false vacuum which expands and then contracts. Relativistic effects,
however, lead to the qualitatively different behavior exhibited by the
other solutions in Fig.~\ref{confdiags1} and Fig.~\ref{confdiags2}. For instance, the evolution
of the IV-III-II/III (solution 2) solution is qualitatively similar to the
IV-I-II/III solution, but is so massive that its evolution is always
hidden behind the black hole event horizon. The IV-I-II'/III-II (solutions
3 and 4) solution is a bubble which has enough kinetic energy to escape
collapse by expanding through the cosmological horizon; observers
inside (or who travel inside from region I) the false vacuum region
will find themselves in an inflationary universe at late times. In the
time-reverse of this solution, a bubble implodes from infinity into
the black hole horizon, and the interior undergoes collapse.

Note that there are two options for the IV-I-II'/III-II solution. Solution 3 corresponds to the situation $A + 3 < B < 3 (A - 1)$, which contains both $\beta_{\rm sds}$ and $\beta_{\rm ds}$ sign changes (see Fig.~\ref{apspotential}). This causes the bubble wall to accelerate in the direction of the true vacuum from both the interior and exterior perspectives. Solution 4 corresponds to the situation $B < A + 3$, which does not contain a $\beta_{\rm ds}$ sign change (see Fig.~\ref{A2_9B3}). This causes the bubble wall to accelerate in the direction of the {\em false} vacuum from the interior perspective. The reason for the disparity is that solution 4 exists only when the interior and exterior cosmological constants are similar in magnitude, and so the wall tension can have greater influence on the dynamics.

The remaining solutions, shown in Fig.~\ref{confdiags2}, have interiors which approach an inflationary
universe at late times, but lie on the opposite side of the wormhole
in the exterior SdS spacetime. To an observer in region III of the SdS
diagram, the IV''-III-II''/IV-I-II (solutions 5) and
IV''-III-II''/IV-III-II solutions (solutions 6 and 7) would appear as a ``sky''
of false vacuum that encroaches from infinity, reaches a minimum
radius and then expands back out. It has been pointed out by Bousso
\cite{Bousso:2004tv} that at late times (in an asymptotically flat
spacetime) the wall trajectory according to the observer in region III
of the S(dS) diagram approaches that of a true vacuum
bubble~\cite{C77,CC77,CD80}. Amusingly, the SdS observer will think he
is in a true vacuum bubble surrounded by a large region of false
vacuum, while the dS observer will think she is in a false vacuum
bubble surrounded by a large region of true vacuum.

This symmetry between true and false vacuum bubbles is made manifest
in the analysis of the Coleman-De Luccia~\cite{CD80} instanton, which
describes the production of both true and false vacuum bubbles
\cite{LW87}.  These are zero energy solutions, and so we should look
for an $M=0$ unbound solution; this corresponds (via Eq.~\ref{Qtom})
to $Q\rightarrow -\infty$, and we see from Fig.~\ref{apspotential}, \ref{gfpotential}, \ref{A1B2}, and \ref{A2_9B3} that the IV''-III-II''/IV-I-II solution (solution 5) or the IV''-III-II''/IV-III-II solution (solution 6), depending on the values of $\Lambda_{+}$ and $\Lambda_{-}$, can be identified as the analytically continued false vacuum instanton.  The radius of the
bubble wall at the turning point is found by considering the limit as the
potential (Eq.~\ref{potential}) goes to $-\infty$, where on the right
(unbound) side of the potential hump the $z^2$ term dominates. Solving
for $r$ using Eq.~\ref{ztor}, we find the radius at turnaround to be
\begin{equation}\label{r_0inst}
r = 6k \left[\left|\left(\Lambda_{+} + \Lambda_{-} +3k^2 \right)^2 - 4\Lambda_{+}\Lambda_{-} \right| \right]^{-1/2},
\end{equation} 
which agrees with the previous literature \cite{BKT87} (see also
\cite{G94}). Since the Schwarzschild mass is zero, we are now matching
two pure de Sitter spacetimes across the bubble wall. The conformal
diagram for the exterior dS region (right) only contains the area
between the vertical dashed lines (which are now identified as $r=0$
surfaces) in solution 5 and 6 of Fig.~\ref{confdiags2}. The interior half (left)
of the diagram remains unchanged. It can be seen that at turnaround,
the bubble will be larger than both the interior horizon size and the
(exterior) horizon size of the region it has replaced.

In the IV-III-II''/III-II solution (solutions 12 and 13), the region of false
vacuum surrounding the observer in region III of the SdS diagram would
begin very small and then expand out of the cosmological horizon. This
solution will also have a time-reversed version in which the
surrounding region of false vacuum implodes. For the
IV''-III''-II''/IV-III-II solution (solutions 8 and 9), the ``sky'' of false
vacuum would forever reside outside of the horizon of a region III
observer. Finally, the N/III-II solution (solutions 10 and 11), which we will
define as solutions with mass greater than or equal to the Nariai
mass, will be an exploding (or imploding in the time-reversed
solution) bubble of false vacuum centered on $r=0$.

In a series of papers, Farhi et. al. \cite{GF87,FGG90} discussed the
application of the Penrose theorem \cite{P65} to the one-bubble
spacetimes discussed above. If the null energy condition (NEC) holds
(as it does for the postulated energy momentum tensor) and there
exists a non-compact Cauchy surface (as in the full SdS spacetime),
then the existence of a closed anti-trapped surface in the spacetime
implies the presence of an initial singularity. Since each point on
the conformal diagrams in Fig.~\ref{confdiags1} and Fig.~\ref{confdiags2} represents a
two-sphere, such an anti-trapped surface exists if the ingoing and
outgoing future-directed null rays both diverge. For example, the
2-sphere represented by point $P_1$ shown in Fig.~\ref{confdiags1},
solution 1, is a closed anti-trapped surface. This can be seen by
following the null ray (null rays are denoted by the dashed lines in
Fig.~\ref{confdiags1}) from $r=0$ in region III of the dS diagram to
$P_1$ and noting that $r$ increases monotonically as $P_1$ is
approached (the null rays are diverging). But following the
future-directed null rays in the opposite direction from $r=0$ in
region IV of the SdS diagram, across the bubble wall, and into the
false vacuum region, shows that they also diverge. Thus, an initial
singularity is necessary for this solution to exist at and near
$P_1$. This spacetime also, however, contains regions without
anti-trapped surfaces. Following the future-directed null rays to
point $P_2$, for example, we see that the ingoing rays diverge, but
the outgoing rays converge. For examples, see solutions 1-4 in
Fig.~\ref{confdiags1}, where regions which contain anti-trapped surfaces
are shaded blue (dark) and the regions which do not are shaded green
(light).

If we cut the IV-I-II/III solution (solution 1) in the expanding phase on
a spacelike hypersurface at a time where the radius of the bubble wall
satisfies $r>r_{BH}$, then the spacetime would not necessarily contain
an initial singularity~\footnote{There are anti-trapped surfaces in
region II' of the SdS diagram, and there is a noncompact Cauchy
surface $C$ for it, so the Penrose theorem applies, but only indicates
that geodesics are incomplete in region II' because they reach its
edge (the past Cauchy horizon of $C$.) The region is thus extendible
(into regions I, III' and IV') rather than singular. Full dS has only
compact Cauchy surfaces so the theorem does not apply.}. We can remove
the initial singularity from the IV-I-II'/III-II (solution 3 and 4) solution as
well by cutting on the same surface. Both the IV-I-II/III (solution 1) and
IV-I-II'/III-II (solution 3 and 4) solutions are therefore classically
buildable. The IV-I-II'/III-II solution (solutions 3 and 4) is the only example
where it is possible to form an inflationary universe from classically
buildable initial conditions, but only exists when the interior and
exterior cosmological constant are almost equal ($B<3(A-1)$). This
solution might be of interest in understanding transitions between
nearly degenerate vacua, for example in the context of eternal
inflation.

\begin{figure}[h!]
\epsfig{file=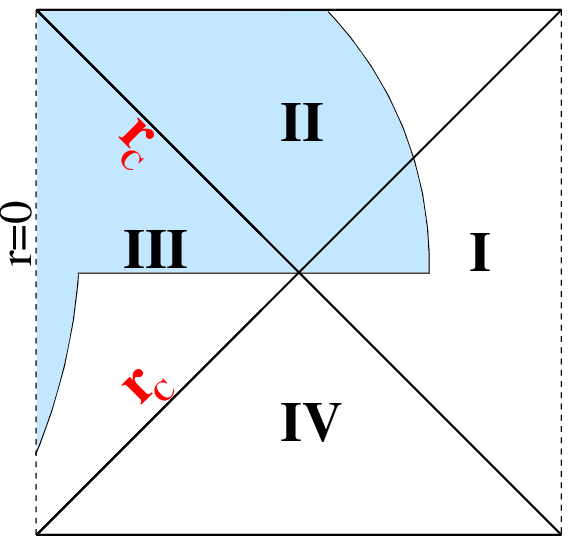,height=2.1cm}
\epsfig{file=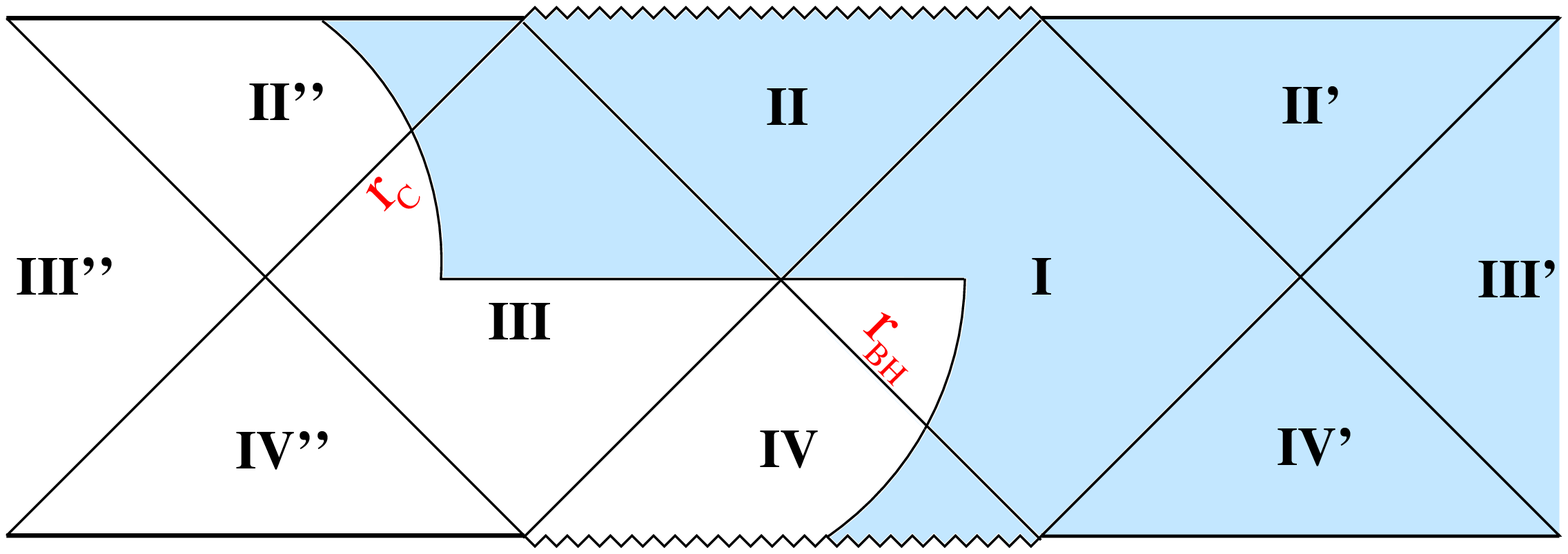,height=2.1cm}
\caption{Tunneling spacetime \label{tunnelst}}
\end{figure}

Given the existence of a classically forbidden region in
Fig.~\ref{apspotential},~\ref{gfpotential},~\ref{A1B2}, and ~\ref{A2_9B3}, one might ask if
tunneling is allowed from one of the recollapsing solutions (solutions 1
and 2) to one of the expanding solutions (solutions 5-9 ) on the other
side of the potential hump. This event, shown in Fig.~\ref{tunnelst},
would constitute a violation of the NEC, and so the Penrose theorem
would no longer apply to the antitrapped surfaces that exist after the
tunneling event. Such a process is indeed apparently
allowed~\cite{FGG90,FMP89}, and would describe the quantum creation of
an inflationary universe from classically buildable initial conditions
(if the initial condition is the IV-I-II/III solution in solution 1).

This is a rather strange transition, however, as the unbound solution
is behind the wormhole in SdS (see Fig.~\ref{tunnelst}). An observer
in region I would see the bubble expand, reach its turning point, and
then disappear, only to be replaced by a black hole. An observer
inside the bubble would see the wall expanding away and -- just as it
is about to turn around and start collapsing -- instead disappear
behind the cosmological horizon. This observer will be inside an
inflationary universe, but forever disconnected from region I. If the
black hole in the SdS spacetime then evaporates, the baby universe
will become completely topologically disconnected.

Having considered one circumstance in which an NEC violation
precipitates the creation of an inflationary universe, one might ask
if there are others. Quantum fluctuations of a scalar field in de
Sitter space can violate the NEC, and so one might imagine that any of
the solutions that we have discussed which allow inflation inside the
bubble could be spontaneously created somewhere along their
trajectories. One example of this direct production of baby universes
is a fluctuation into one of the unbound solutions (solutions 5-9). Such
scenarios have been considered in the context of the stochastic
approach to baby universe production by Linde~\cite{L91} and in
reference to eternal inflation by Carroll and
Chen~\cite{Carroll:2004pn}. 

\begin{figure}[h!]
\includegraphics[height=4.2cm]{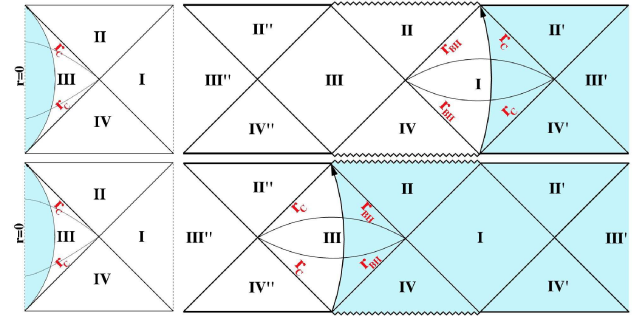}
\caption{Conformal diagrams for the thermal decay of de Sitter vacua. The upper diagram corresponds to $B < 3 (A-1)$, the lower diagram corresponds to $B > 3 (A-1)$. \label{thermalons}}
\end{figure}

Another example of the direct fluctuation into an inflationary universe is the thermal decay of de Sitter vacua discussed by Garriga and Megevand~\cite{GM04} (Ref.~\cite{GHTW04} discusses a related mechanism). This process consists of the fluctuation from  empty de Sitter of a bubble in unstable equilibrium between expansion and collapse. This static solution is identified as the set of spacetimes which sit on top of the potentials in Fig.~\ref{apspotential}, \ref{gfpotential}, \ref{A1B2}, and \ref{A2_9B3}. The two possible configurations are shown in Fig.~\ref{thermalons}, where it can be seen that the bubble wall can lie on either side of the worm hole depending on the sign of $\beta_{\rm sds}$ at the top of the potential. It can be shown that the ``Nariai limit" in Ref.~\cite{GM04} corresponds to $B = 3 (A -1)$, where the $\beta_{\rm sds}$ sign change occurs at the max of the potential.

With these considerations, there are at least three competing channels
for the formation of baby universes: direct production, the FGG
mechanism, and the zero-mass false vacuum instanton. It would be
desirable to develop a scheme to directly compare the relative
probabilities of each of these processes, and this is currently being explored.

Our full catalog of solutions is also interesting in regards to a
recent proposal~\cite{VT99,DV05} that false vacuum regions, assumed to
be larger than the interior horizon, must at all times be larger than
the exterior, true vacuum, horizon. The basis of this conjecture is
the condition that the divergence of a congruence of future directed
null geodesics (defined as $\theta$) must satisfy
\begin{equation}\label{dthetadT}
\frac{d\theta}{dT} \leq 0,
\end{equation}
 where $T$ is an affine parameter, if the NEC holds for all $T$. Null
rays in the dS and SdS spacetimes satisfy this inequality (in dS, the
inequality is exactly zero), but we should check that the junction
conditions do not violate it. One requirement imposed by
Eq.~\ref{dthetadT} is that the divergence of the null rays does not
increase at the position of the wall as they go from a true vacuum
region into a false vacuum one. Along any given null geodesic in the
bubble interior or exterior, the value of $r$ is either increasing or
decreasing monotonically as a function of $T$. We can therefore state
the condition Eq.~\ref{dthetadT} as: one cannot have a null ray along
which $dr/dT \leq 0$ outside the bubble and $dr/dT \geq 0$ inside the
bubble. Surveying the solutions in Fig.~\ref{confdiags1} and Fig.~\ref{confdiags2}, we see that this is indeed always true.

The authors of Ref.~\cite{VT99,DV05} intended to demonstrate that if
one requires the false vacuum region to be larger than the interior
horizon size at all times (so that inflation is unstoppable), it is
necessarily larger than the exterior horizon size. Although all of the
allowed one-bubble spacetimes satisfy the condition
Eq.~\ref{dthetadT}, there are only two examples in which all observers
agree that this requirement is met: the false vacuum instanton
(IV''-III-II''/IV-I-II solution (solution 5), with $M=0$) and the
IV-I-II'/III-II solution (solution 3 and 4, after turn-around). In every other
case (including the FGG spacetime in Fig.~\ref{tunnelst}), the
observers in region I of the SdS conformal diagram will see only a
{\em black hole} horizon sized volume replaced by the false vacuum
bubble. We therefore conjecture that if one requires the false vacuum
region to be larger than the interior horizon size at all times, then
it will replace a volume larger than the exterior horizon size
according to {\em only some} observers. If one relaxes this
requirement, then there are a diverse range of solutions which might
describe the spawning of an inflationary universe. For example, the
bubbles in solutions 3, 4 and 10-13 of Fig.~\ref{confdiags1} and~\ref{confdiags2} grow from an arbitrarily small size.
 
Having discussed the character of the various solutions, we turn now
to a potentially dangerous detail, which is particularly important for
the FGG mechanism: there exists a classical instability against
aspherical perturbations in the spherically symmetric solutions to the
junction conditions.

\section{Perturbations}\label{perts}

The solutions described in Sec.~\ref{sols} assume that the region of
false vacuum is spherically symmetric. The stability of these
solutions against aspherical perturbations has important consequences,
especially if one hopes to build plausible cosmologies. That there
might be an instability in domain walls was first discussed by Adams,
Freese and Widrow~\cite{AFW91}. The bubble wall can trade volume
energy for surface energy and wall kinetic energy locally as well as
globally, and so the bubble wall will become distorted if different
sections of the wall have different kinetic energies.  As long as the
local distortions of the wall remain small compared to the size of the
background solution's radius, this process can be formulated
quantitatively as perturbation theory around a background spherically
symmetric solution.

Previous authors~\cite{AFW91,GV91,G93} have considered perturbations
on expanding bubbles of true-vacuum~\cite{C77,CC77,CD80}, which have
zero total energy (surface, volume, and kinetic energies canceling),
and so can expand asymptotically. As was first pointed out by Garriga
and Vilenkin~\cite{GV91}, even though local observers on the bubble
wall see perturbations grow, external observers see them freeze out
because they do not grow faster than bubble radius.

The story is different for the bound (solutions 1 and 2) and unbound
(solutions 5-9) false vacuum bubbles: since they reach a turning point,
the perturbations have a chance to catch up to the bubble's expansion
and become nonlinear. This also presumably has implications for the thermal decay mechanism of Garriga and Megevand~\cite{GM04}, depending on the duration of time the bubble wall sits in unstable equilibrium between expansion and collapse (see discussion in Sec.~\ref{sols}). The remainder of this work will focus on the
instability of the bound IV-I-II/III solution (solution 1), since
physically plausible initial conditions may be clearly
formulated. There is no obvious set of initial conditions for the
perturbations on the unbound solutions, and so we simply observe that
the results we will obtain for the bound solutions apply qualitatively
here as well.

To simplify the problem, we assume that the full gravitational problem
described in the previous sections can be treated as motion of the bubble
wall in a fixed SdS background. This assumption must be validated (as
we do below), but we are mainly interested in the low-mass bound
solutions for which we might expect the gravitational contributions to
be small.  Assuming that a thin spherically symmetric bubble wall
separates an internal dS from an external SdS spacetime, we can employ
the action~\cite{G93,AFW91,GV91}:

\begin{equation}\label{matteraction}
S=-\sigma \int d^{3}\xi \sqrt{-\gamma}+\epsilon \int d^{4}x \sqrt{-g}, 
\end{equation}
where $\sigma$ is the surface energy density on the bubble wall,
$\gamma_{ab}$ ($a,b=1,2,3$) is the metric on the worldsheet of the
bubble wall, $\epsilon$ is the difference in volume energy density on
either side of the bubble wall:
\begin{equation}
\epsilon = \frac{\Lambda_{+}-\Lambda_{-}}{8 \pi},
\end{equation}
 and $g_{\alpha \beta}$ is the metric of the background spacetime.

\subsection{Wall Equation of Motion}\label{walleom}
The equation of motion resulting from Eq.~\ref{matteraction} is~\cite{G93}:
\begin{equation}\label{simpeom}
g^{a b}K_{a b}=-\frac{\epsilon}{\sigma},
\end{equation} 
where $K_{a b}$ is the extrinsic curvature tensor of the worldsheet of
the bubble wall,
\begin{equation}
K_{a b}=-\partial_{a}x^{\mu} \partial_{b}x^{\nu} D_{\nu}n_{\mu}, 
\end{equation} 
and $D_{\nu}$ is the covariant derivative and $n_{\mu}$ is the unit
normal to the bubble wall worldsheet.

We will use the static foliation of the SdS spacetime (see
Eq.~\ref{gsds}) as the coordinates $x^\mu$ for the background
spacetime. The world sheet is given coordinates $(\tau,\theta,\phi)$
as in Eq.~\ref{wallmetric}, and has metric:
\begin{equation}
\gamma_{a b}=g_{\mu \nu}\partial_{a}x^{\mu}\partial_{b}x^{\nu},
\end{equation}
with the gauge freedom in choosing $\tau$ fixed by
\begin{equation}
\frac{dt}{d\tau}\equiv t'=\frac{\sqrt{a+r'^2}}{a},
\end{equation}
so that $\gamma_{\tau\tau}=-1$. Here and henceforth primes will
denote derivatives with respect to $\tau$. The other non-zero
components of $\gamma_{a b}$ are $\gamma_{\theta \theta} = r^2$ and
$\gamma_{\phi \phi} = r^2 \sin^2 \theta$.

The first task at hand is to find the worldsheet's unit normal, which
by spherical symmetry has only $r$ and $t$ components.  Requiring
orthogonality to the worldsheet ($g_{\mu\nu} n^{\nu} \partial_{a}
x^{\mu}=0$) and unit norm ($g_{\mu\nu}n^{\mu}n^{\nu}=1$) yields its
components:
\begin{subequations}\label{sdsn}
\begin{equation}
n_{t}=-r',\ \ n_{r}=t'.
\end{equation}
\end{subequations}
The components of $K_{ab}$ are given by
\begin{subequations}\label{kab}
\begin{equation}
K_{\tau\tau} = \left[ r'' + \frac{1}{2}\frac{da}{dr}\right] (a + r'^2)^{-1/2},
\end{equation}
\begin{equation}
K_{\phi \phi} = -r a t' \sin^2 \theta = K_{\theta \theta} \sin^2 \theta.
\end{equation}
\end{subequations}

Substituting Eq.~\ref{kab} into Eq.~\ref{simpeom} gives the equation of motion for the bubble wall:
\begin{equation}\label{sdseom}
r''=\frac{\epsilon}{\sigma} \sqrt{a+r'^2} - \frac{2}{r} (a+r'^2) - \frac{1}{2} \frac{da}{dr}.
\end{equation}
 
Eq.~\ref{potential} supplies the velocity of the bubble at some position along its trajectory
\begin{equation}\label{v_0}
z'=[Q-V(z_{0})]^{1/2}.
\end{equation}
Choosing this boundary condition is effectively restricting ourselves
to the IV-I-II/III (solution 1) or IV-III-II/III (solution 2) solutions. Since
the solutions to Eq.~\ref{sdseom} approximate the dynamics of the
junction condition problem, we should parametrize by $A$, $B$, and
$Q$. This can be done by using the conversions defined in
Sec.~\ref{junctioneom}, and gives:
\begin{eqnarray}
\label{dimwall}
z'' &=& -\frac{3(B-A)}{c}\sqrt{a(-Q)+z'^2} \\ \nonumber
&-& \frac{2}{z} (a(-Q)+z'^2) - \frac{(-Q)}{2}\frac{da}{dz},
\end{eqnarray}
where $a$ is written in terms of $z$ as
\begin{equation}
a=1-\frac{12}{c z (-Q)} - \frac{12 A}{c^2 (-Q)} z^2,
\end{equation}
and 
\begin{equation}
c \equiv \left[|(A+B+3)^{2}-4AB|\right]^{\frac{1}{2}}.
\end{equation}

To justify the use of the simplified dynamics described above,
Eq.~\ref{dimwall} was numerically integrated, and the position of the
turning point compared to the corresponding point on the full junction
condition potential. Over the range of $Q$ corresponding to the bound
solutions, we find excellent quantitative agreement (well within 1\%)
between the turning points of the solutions to Eq.~\ref{dimwall} and
the junction condition potential. This was repeated with equally good
results for the weak, GUT, and Planck scale potentials and also for
various initial positions between the black hole radius and the
potential wall (turning point). This shows that to zeroth order,
dynamics as motion in a background is valid, and strongly suggests
that it will be at higher orders as well.

\subsection{Perturbations \label{perturbations}}

We are now in a position to discuss the first-order perturbations on
the spherically-symmetric background solutions discussed in
Sec.~\ref{walleom}. Physical perturbations are normal to the
worldsheet of the (background) bubble wall, and can be described by
scalar field $\phi(x)$ by taking the position of the perturbed
worldsheet to be
\begin{equation}
\bar{x}^{\mu}=x^{\mu} + \phi(x) n^{\mu},
\end{equation}
where $x^{\mu}$ is the spherically symmetric solution, and $n^{\mu}$
is the unit normal to the worldsheet. It is assumed that $\phi$ is
much smaller than the bubble wall radius, so that a perturbative
analysis can be made.
 
 The equation of motion for the perturbation field $\phi(x)$ in a
curved spacetime background can be derived from the action Eq.~\ref{matteraction} after expanding to second order in $\phi(x)$
\cite{G93}
\begin{equation} \label{meom}
\bigtriangleup \phi - \left[-R_{\mu \nu} h^{\mu \nu} + R^{(3)} - \frac{\epsilon^2}{\sigma^2} \right] \phi = 0,
\end{equation} 
where 
\begin{equation}
\bigtriangleup \phi=\frac{1}{\sqrt{-\gamma}} \partial_{a} \left(\sqrt{-\gamma} \gamma^{a b} \partial_{b} \phi \right),
\end{equation}
and $h^{\mu \nu}$ is 
\begin{equation}
h^{\mu\nu}=g^{\mu\nu}-n^{\mu}n^{\nu}.
\end{equation}

To solve the equation of motion, we can decompose $\phi(x)$ into spherical harmonics
\begin{equation}\label{phimodes}
\phi(x) = \sum_{l,m} \phi_{lm}(\tau) Y_{lm}(\theta,\phi),  
\end{equation}
and separate variables to get an equation for $\phi_{lm}(\tau)$. The
geometrical factors in Eq.~\ref{meom} become dependent on
$\theta$ or $\phi$ only at second order, so we will always be
able to make this decomposition.

$\bigtriangleup \phi$ is then given by:
\begin{equation}\label{triangle}
\bigtriangleup \phi_{lm} = -\left(\partial_{\tau}^{2} + \frac{2 r'}{r} \partial_{\tau} + \frac{l(l+1)}{r^2} \right) \phi_{lm}.
\end{equation}
The components of $h^{\mu\nu}$ are:
\begin{subequations}\label{h}
\begin{eqnarray}
&& h^{tt}=-\frac{a+r'^2}{a^2},\ \  h^{rr}=-r'^2, \\
&& h^{\theta\theta}=h^{\phi\phi} \sin^2\theta =\frac{1}{r^2}.
\end{eqnarray}
\end{subequations}
The components of the Ricci tensor are given by:
\begin{subequations}\label{Rt}
\begin{equation}
R_{tt}=\frac{a}{2}\partial_{r}^2 a + \frac{a}{r} \partial_{r} a= - a^2 R_{rr},
\end{equation}
\begin{equation}
R_{\phi\phi} = R_{\theta\theta} \sin^2 \theta = (1 - a - r \partial_{r} a).
\end{equation}
\end{subequations}
Contracting equations \ref{h} and \ref{Rt} gives:
\begin{equation}\label{hR}
R_{\mu\nu}h^{\mu\nu}=\frac{2(1-a)}{r^2} - \frac{3 \partial_{r}a}{r} - \frac{\partial_{r}^2 a}{2}=3 \Lambda_{+}.
\end{equation}
The Ricci scalar on the world sheet is
\begin{equation}
\label{R}
R^{(3)}=\frac{2}{r^2} (1 + r'^2 + 2r r''),
\end{equation}
where $r''$ is given by Eq.~\ref{sdseom}.

After substituting Eq.~\ref{triangle}, Eq.~\ref{hR}, and Eq.~\ref{R} into
Eq.~\ref{meom}, the equation of motion for $\phi_{lm}(\tau)$ is
\begin{widetext}
\begin{equation}
\label{sdsddphi}
\phi_{lm}''=\left[\frac{\epsilon^2}{\sigma^2} - \frac{4
\epsilon}{\sigma r} \left(a+r'^2 \right)^{1/2} + 3 \Lambda_{+} +
\frac{2}{r}\frac{da}{dr} + \frac{6 r'^2}{r^2} - \frac{2(1-4a)}{r^2} -
\frac{l(l+1)}{r^2} \right] \phi_{lm} - \frac{2r'\phi_{lm'}}{r}.
\end{equation}
In terms of the dimensionless variables of the junction condition
problem this reads:
\begin{eqnarray}
\label{sdsdimeom}
&&\Phi_{lm}'' = -\frac{2 z'}{z}\Phi_{lm}' \\ &&+ \left\{\frac{108
A}{c^2} + \frac{9 (A-B)^2}{c^2} + \frac{12 (B-A)}{c
z} \left( a(-Q)+z'^2 \right)^{1/2} + \frac{2 (-Q)}{z^2}(4a-1) + \frac{6 z'^2}{z^2} +
\frac{2 (-Q)}{z} \frac{da}{dz} - \frac{l(l+1)(-Q)}{z^2}
\right\}\Phi_{lm}, \nonumber
\end{eqnarray}
\end{widetext}
where $\Phi$ is the dimensionless perturbation field defined similarly
to $z$ (see Eq.~\ref{ztor}). The first term acts as a (anti)drag on
(shrinking) growing perturbations. The last term in this equation is
always negative, acting as a restoring force. Perturbations will grow
when the other terms (which are positive over most of the trajectory
in the expanding phase) in this equation dominate. Further, the last
term indicates that lower $l$ modes will experience the largest
growth. The full details of the solutions, however, require a
numerical approach, to which we now turn.

\section{Application to the Farhi-Guth-Guven mechanism}
\label{sec-results}

The possibility of creating an inflating false-vacuum region via
quantum tunneling (described in Sec.~\ref{sols}) has been investigated
only under the assumption of spherical symmetry, and this would be
grossly violated if perturbations on the bubble wall become
nonlinear. In this section, we investigate the circumstances under
which this is the case. The two basic questions at issue are: first,
when do perturbations go nonlinear for some given set of initial
perturbations, and second, what initial perturbations can be expected.

\subsection{Dynamics of the Perturbation Field}\label{dynamics}

Let us begin with the first issue. Since Eq.~\ref{sdsdimeom} is a
second order ODE, it can be decomposed into the sum of two linearly
independent solutions
\begin{eqnarray}\label{levolve}
\Phi_{\rm lm}(T)&=&\Phi_{\rm lm}(T=0) f(l,z_{0},Q,T) \nonumber \\
 && + \Phi_{\rm lm}'(T=0) g(l,z_{0},Q,T).
\end{eqnarray}
The functions $f(l,z_{0},Q,T)$ and $g(l,z_{0},Q,T)$ can be
found numerically by alternately setting $\Phi_{\rm lm}(T=0)$ and
${\Phi}_{\rm lm}'(T=0)$ to zero, then evolving the coupled
Eq.~\ref{sdsdimeom} and Eq.~\ref{dimwall} numerically for a time
$T$ with initial conditions for $Q$, $z_{0}$, and $l$. If the
bubble is to tunnel, it will do so at the time $T_{\rm max}$, when
the bubble reaches its maximum radius and begins to re-collapse.
Given $f$ and $g$ at time $T_{\rm max}$, the size of the
perturbations at the turning point for any $z_{0}$, $Q$, $l$,
$\Phi_{\rm lm}(T=0)$, and $\Phi_{\rm lm}'(T=0)$ can be
determined.  An RK4 algorithm with adaptive step size was used to
solve for $f$ and $g$, with numerical errors well
within the $1\%$ level.

\begin{figure*}
\includegraphics[width=8.6cm]{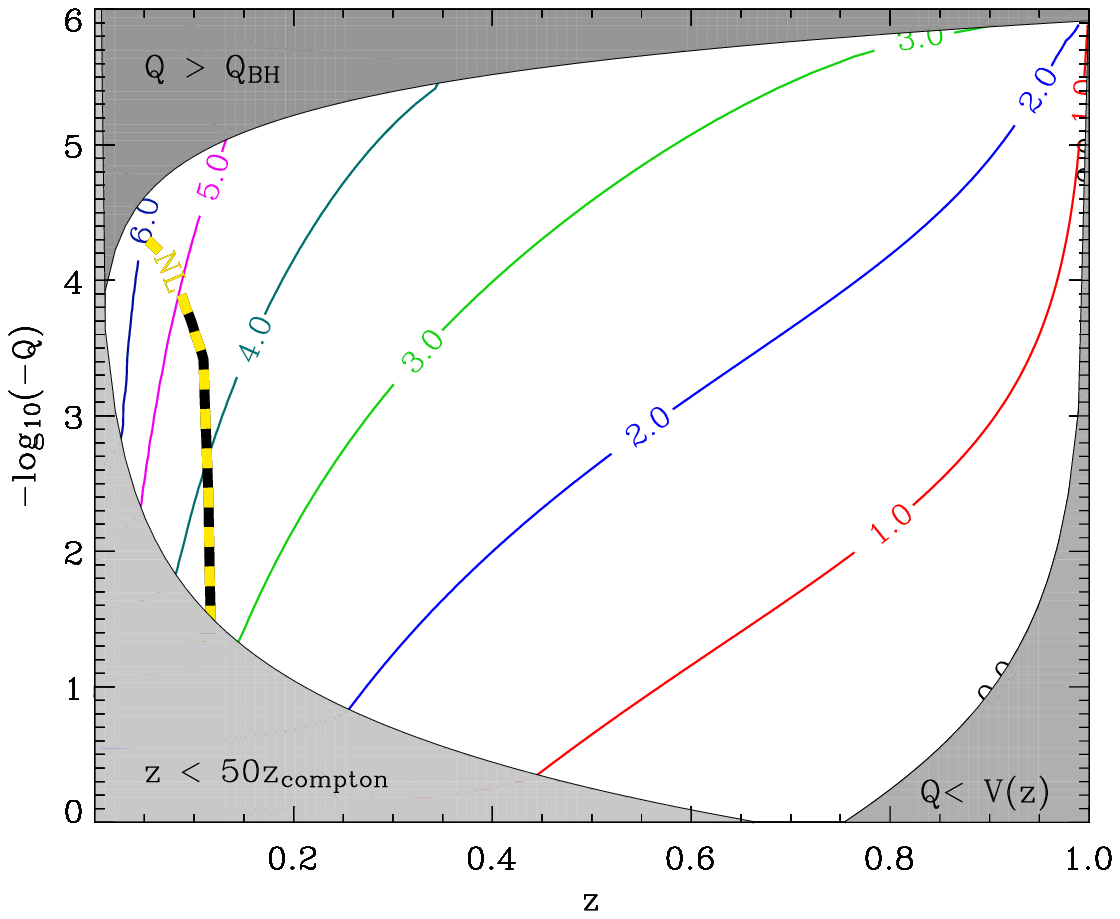}
\includegraphics[width=8.6cm]{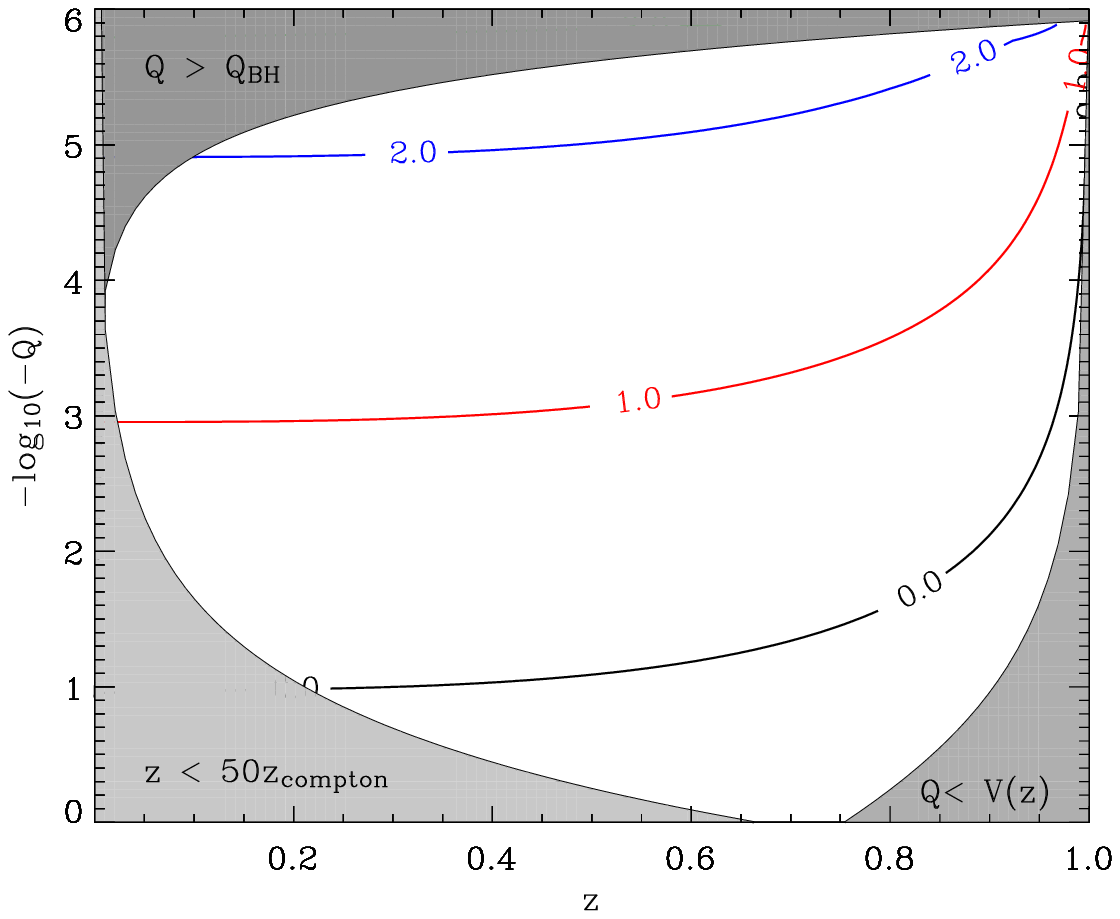}
\includegraphics[width=8.6cm]{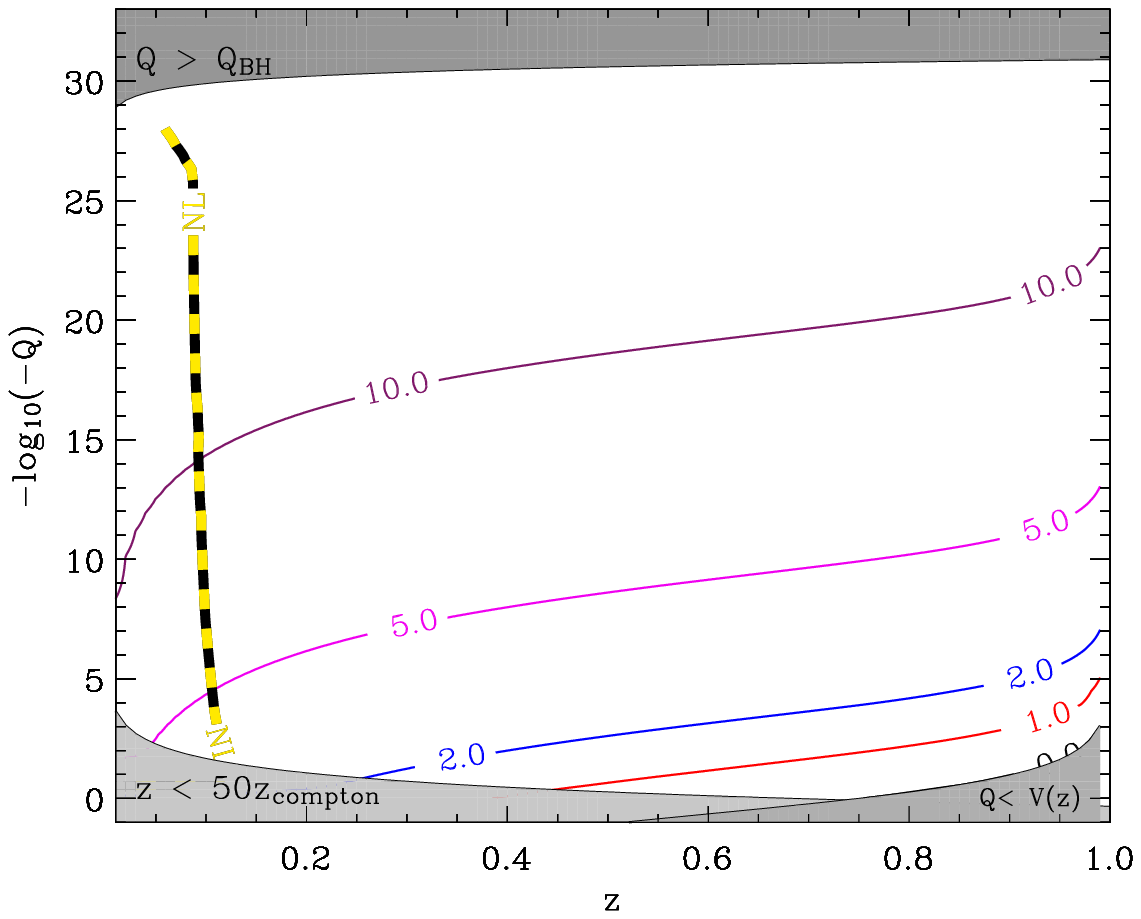}
\includegraphics[width=8.6cm]{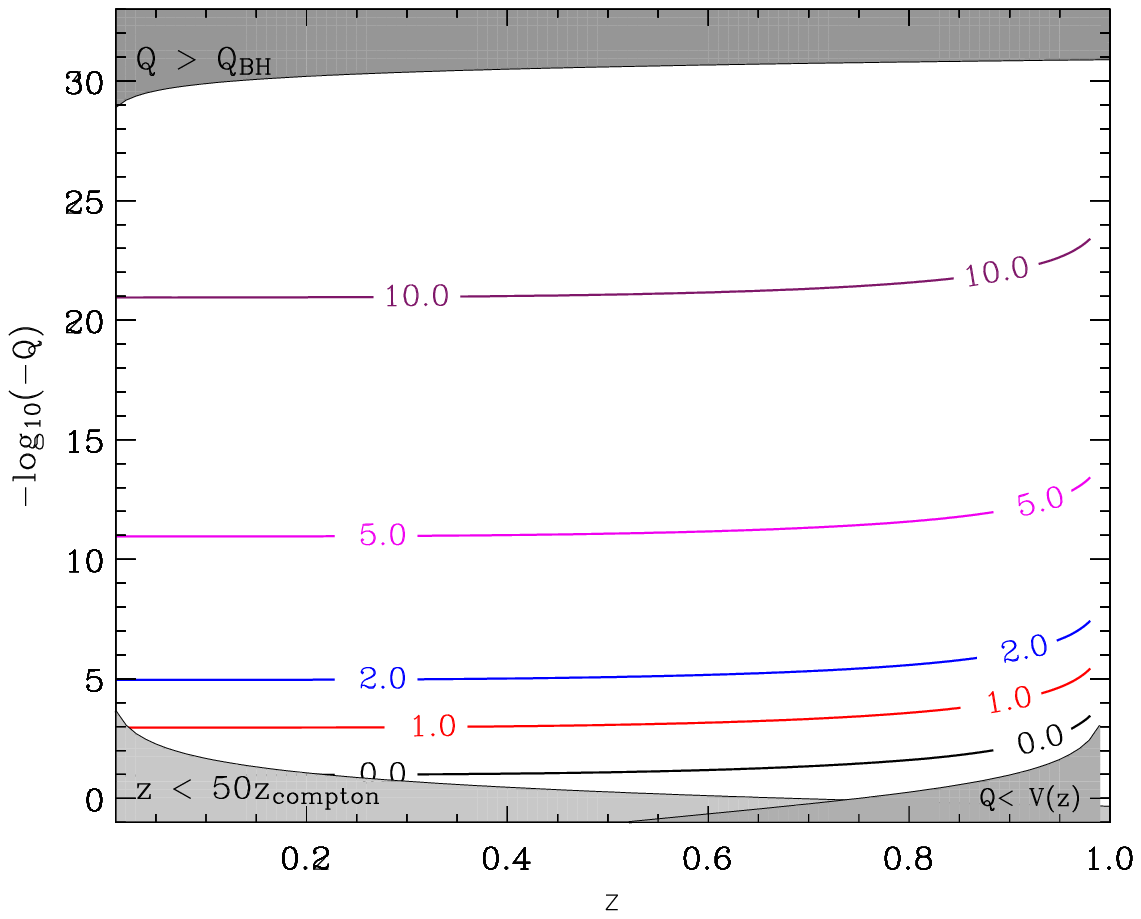}
\caption{(Color online) Contour plot of $Log_{10}[f(l=1,z_{0},Q,T_{\rm max})]$ (left) and  $Log_{10}[g(l=1,z_{0},Q,T_{\rm max})]$ (right) for $M_I=10^{14}$ GeV (top) and $M_I=100$ GeV (bottom). \label{fg}}
\end{figure*}

The results of this analysis for $l=1$ and for the low (weak) and
intermediate (GUT) inflation scales discussed below Eq.~\ref{scaleAB}
are shown in Fig.~\ref{fg}. The solid lines show contours of constant (log)
amplification factor $f$ (left) and $g$ (right) versus the bubble
starting radius $z_0$ and mass parameter $Q$, with bubble mass
increasing toward the top. The shaded regions indicate regions which
we have disallowed as bubble starting radii because the bubble would
not be classically buildable for $r<r_{\rm BH}$ (marked as $Q > Q_{\rm
  BH}$), or the bubble is in the forbidden region $Q < V(z)$ of the
effective 1D equation of motion Eq.~\ref{juncteom}, or the bubble
would be too small to be treated classically. We choose the latter
radius as fifty times the Compton wavelength $z_{\rm compton}$ of a
piece of the bubble wall~\footnote{The mass of a piece of wall of
  scale $s$ is $M \simeq s^2 \sigma$, where $\sigma$ is the wall
  surface energy density; the Compton wavelength is then found by
  setting $M=s^{-1}$, yielding $s = z_{\rm compton} \simeq
  \sigma^{-1/3}$.}. The choice of fifty Compton wavelengths is rather
arbitrary; the effect of a larger bound would be to exclude more of
the parameter space in Fig.~\ref{fg}.  This (unshaded) parameter space
includes all classical initial conditions which could be set up by the
observer in region I of the SdS conformal diagram.

It can be seen in Fig.~\ref{fg} that the growth of the perturbations
is in general larger for higher-mass bubbles (smaller $|Q|$, larger
$-\log_{10}(-Q)$). The lower the inflation scale, the closer to zero
the peak in the potential function becomes, and the smaller $|Q|$
(higher mass) bubbles are allowed, so at low inflation scales $f$ and
$g$ can be very large. Growth for the Planck-scale inflation bubbles
is very small, with $f$ of order 10 and $g$ of order 1, and is not
plotted.

\begin{figure}
\includegraphics[width=8.6cm]{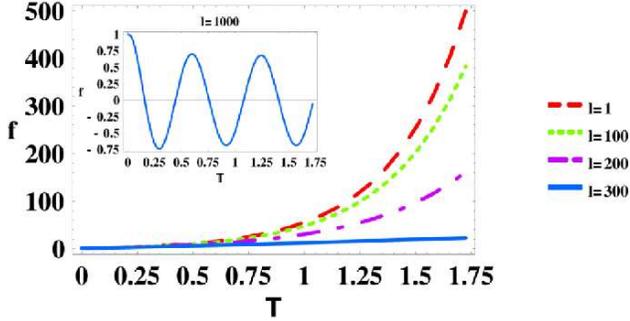}
\caption{$f(l,z_{0}=.5,Q=-10^{-4},T)$ for various $l$. The inset
  shows the oscillatory behavior of $f$ for large $l$. \label{usj}}
\end{figure}

The enhanced growth at small $|Q|$ is due to the suppression of the
term in Eq.~\ref{sdsdimeom} proportional to $l(l+1)(-Q)$, which always
acts to stabilize the perturbations. Another consequence of this
suppression is that the range in $l$ over which solutions are unstable
depends on $Q$; as a general rule of thumb, approximately a few times
$(-Q)^{-1/2}$ $l$-modes are unstable (note that this is unlike the
case true vacuum bubbles, for which only the $l=0,1$ modes are
unstable).  An example of the $f$ function for $Q=-10^{-4}$ with the
intermediate (GUT) inflation scale parameters is shown in
Fig.~\ref{usj}. The $f$ functions for very large $l$ modes are stable
and approach sinusoids with amplitudes less than one (see the inset of
Fig.~\ref{usj}), meaning that the perturbations are never larger than
their initial size.

\subsection{Initial Conditions and Evolution to the Turning Point}

Having fully characterized the growth of the perturbations, we now
require an estimate for their initial values when the bubble is
formed.  There is no reason to expect that a region of false vacuum
will fluctuate into existence with anything near spherical symmetry,
nor is it likely to have thin walls (there is no instanton or other
mechanism to enforce these symmetries). Since low-$l$ (relative to
$(-Q)^{-1/2}$) modes are unstable, an initially aspherical bubble will
only become more aspherical; this is in marked contrast to true vacuum
bubbles, which both start spherical, and tend to become more spherical
as they expand.

Suppose, however, that we consider the best-case scenario in which a
bubble {\em is}, by chance or design, spherically symmetric.  It will
nevertheless inevitably be dressed with zero-point quantum
fluctuations of the perturbation field. We may then check whether
these fluctuations alone, considered as initial values for the
perturbations of a bubble starting with a given $Q$ and $z_{0}$,
suffice to make the bubble nonlinearly aspherical by turnaround.

We assume that the ensemble average of the quantum fluctuations at the
time of nucleation is zero; but the ensemble average of the square of
the field (the space-like two-point function $\langle
\Phi(\theta,\phi) \Phi(\tilde{\theta},\tilde{\phi})\rangle \equiv
\langle \Phi \tilde{\Phi} \rangle$) will not generally vanish.  We can
write the mode functions (Eq.~\ref{phimodes}) in terms of it as:
\begin{eqnarray}\label{philm2}
\langle \Phi_{\rm lm}^{2} \rangle & = & \\
&& \int d\Omega d\tilde{\Omega} \nonumber
 \langle \Phi \tilde{\Phi} \rangle  Y_{\rm lm}(\theta,\phi) Y_{\rm lm}^{*}(\tilde{\theta},\tilde{\phi}).
\end{eqnarray}
By spherical symmetry, the two-point function can be written as a function of the angular separation $\Psi$ between $(\theta,\phi)$ and
$(\tilde{\theta},\tilde{\phi})$, and decomposed into Legendre polynomials:
\begin{equation}
\langle \Phi \tilde{\Phi} \rangle = \sum_{l} C_{l} P_{l}(\cos \Psi). 
\end{equation}
Using the addition theorem for spherical harmonics, we can write this as
\begin{equation}
\langle  \Phi \tilde{\Phi} \rangle  = \sum_{\rm l',m'} \frac{4 \pi}{2l'+1} C_{l
'} Y_{\rm l'm'}^{*}(\theta,\phi) Y_{\rm l'm'}(\tilde{\theta},\tilde{\phi}).
\end{equation}
Substituting this into Eq.~\ref{philm2} and using the orthogonality of the
spherical harmonics yields the relation:
\begin{equation}\label{initcond}
\langle \Phi_{\rm lm}^2\rangle  = \frac{4 \pi C_{l}}{2l+1}.
\end{equation}

Given some space-like two point function at the time the bubble is
nucleated, we can obtain the $C_{l}$ from
\begin{equation}\label{Cl}
C_{l} = \frac{2l+1}{4 \pi} \int_{-1}^{1} d(\cos \Psi) \langle \Phi \tilde{\Phi}\rangle  P_{l}(\cos \Psi)
\end{equation}
and therefore set the typical initial amplitudes of the mode functions
as the r.m.s. value $\langle \Phi_{\rm lm}^2\rangle ^{1/2}$ from
Eq.~\ref{initcond}.  The velocity field can be decomposed into
spherical harmonics just as $\Phi$ was, and the analysis performed
above carries over exactly. The typical initial size of the velocity
mode functions is then given by
\begin{equation}\label{initcondg}
\langle \Phi_{\rm lm}'^2\rangle  = \frac{4 \pi A_{l}}{2l+1}.
\end{equation}
with 
\begin{equation}\label{Al}
A_{l} = \frac{2l+1}{4 \pi} \int_{-1}^{1} d(\cos \Psi) \langle \Phi' \tilde{\Phi}'\rangle  P_{l}(\cos \Psi).
\end{equation}

The initial amplitudes in Eq.~\ref{initcond} and Eq.~\ref{initcondg}
can now be evolved to the turning point, and the mode functions
re-summed. The ensemble average of the r.m.s. fluctuations in $\Phi$
at any time at a given point will then be:
\begin{eqnarray}
\label{Phiavg}
\langle {\Phi(T)}^2\rangle  &=&  \sum_{l} \frac{2l+1}{4 \pi} \langle {\Phi_{\rm lm}(T)}^2 \rangle \\ \nonumber
&=&  \sum_{l} \left[ C_{l}^{1/2} f(l,z_0,q,T) + A_{l}^{1/2} g(l,z_0,q,T) \right]^2, 
\end{eqnarray}
which can be evaluated at $T=T_{\rm max}$.

A full model of the two-point functions $\langle\Phi\tilde\Phi\rangle$
and $\langle\Phi'\tilde\Phi'\rangle$ would involve quantizing the mode
functions on the curved spacetime of the bubble wall worldsheet, which
has a metric depending on $z(T)$.  Further, to treat large
fluctuations, we would need to include non-linear terms in the
equation of motion. The exact two-point function is therefore a rather
formidable object to compute. As a simplified model, we will employ
the two-point functions of a massless scalar field in flat spacetime,
and replace the spatial distance $r$ with the distance along the
bubble wall $r_0\Psi$. This massless scalar corresponds to the
perturbations on a flat wall separating domains of equal energy
density in Minkowski space \cite{GVQ91}. Corrections to this picture
in the presence of curvature should be small over small regions of the
bubble wall. We are also neglecting the large difference in energy
densities across the bubble wall, which will give the field a
(negative) mass to first order. The apparent divergence of the
correlator due to this negative mass will be rendered finite by the
non-linear terms which must be introduced to discuss large
fluctuations. In light of all these difficulties, and several more
approximations we will make below, this should be considered as a
first, rough estimate of the amplitude of the quantum fluctuations on
the bubble at the time of nucleation.

The space-like two-point function in Minkowski space at large separations is given by
\begin{equation}\label{philarger}
\langle \phi({\bf x}) \phi({\bf y}) \rangle =\frac{\sigma^{-1}}{4 \pi r}, 
\end{equation}
where $r \equiv |\bf{x}-\bf{y}|$. As in the work of Garriga and Vilenkin \cite{GVQ91}, we introduce a smeared field operator to obtain a well-defined answer at close separations
\begin{equation}
\phi_{s} \equiv \frac{1}{\pi s^2} \int_{|\bf{y} - \bf{x}|< s} d^2 y \phi(\bf{y}), 
\end{equation}
where $s$ is a smearing length, chosen to be the Compton wavelength
$s=\sigma^{-1/3} $ of
a piece of wall, which is a physically reasonable lower bound on the size of a measurable region. The smeared correlator will then be given by
\begin{eqnarray}\label{smeared2p}
\langle \phi_{s}({\bf x}) \phi_{s}({\bf y}) \rangle = \frac{1}{\pi^2 s^4} && \int_{\rm |{\bf z}-{\bf x}| < s} d^2 z \\ \nonumber && \int_{\rm |{\bf q}-{\bf y}| < s} d^2 q \langle \phi({\bf z}) \phi({\bf q}) \rangle,
\end{eqnarray}
which evaluates at ${\bf x} = {\bf y}$ to
\begin{equation}\label{phismallr}
\langle \phi_{s}^{2} \rangle = (\frac{1}{2} + {\cal G}) \sigma^{-2/3},
\end{equation}
where ${\cal G}=.916...$ is Catalin's constant. We have been unable to
integrate Eq.~\ref{smeared2p} to obtain the exact form of the smeared
two-point function for all $r$. However, it must smoothly interpolate
between the value at $r=0$ in Eq.~\ref{phismallr} to the functional
form at $r \gg s$ given by Eq.~\ref{philarger}. We therefore employ
the following `toy model' smeared field correlator
\begin{eqnarray}\label{phiconnect}
\langle \phi({\bf x}) \phi({\bf y})  \rangle  &=& \frac{\sigma^{-1}}{4\pi s} \left[ (2\pi + 4\pi {\cal G} -1)e^{-r^2/2s^2} \right. \\ \nonumber && \left. + \frac{1}{r/s + 1} \right],
\end{eqnarray}
which has the correct asymptotics. With $r=r_0\Psi$, the dimensionless
form of the toy correlator is given by:
\begin{eqnarray}
\label{eq-2pt}
\langle  \Phi \tilde{\Phi} \rangle  = \frac{(-Q)}{(24 \pi)^2}  && \left[ (2\pi + 4\pi {\cal G} -1)e^{-(R^2/2) \Psi^2} \right. \\ \nonumber && \left. + \frac{1}{R \Psi + 1} \right],
\end{eqnarray} 
where $R\equiv r_0/s$.

The velocity-velocity correlator can be calculated using the
Hamiltonian approach. For spacelike separations, this is given by
\begin{eqnarray}
\langle \phi'( {\bf x}) \phi'( {\bf y}) \rangle &=& \sigma^{-1} \int \frac{|p| d^2 p}{2(2 \pi)^2} \exp[i {\bf p} ({\bf x}-{\bf y} )] \\ \nonumber &=& \sigma^{-1} \int_{0}^{\Lambda} \frac{dp}{4 \pi} p^2 J_{0}(pr)
\end{eqnarray}
where we have introduced a hard momentum cut-off $\Lambda$ to obtain a
finite answer. This cutoff will correspond to the inverse smearing
length, with higher momentum scales higher accounted for in the
smeared operator. The integral can be evaluated in terms of
generalized hypergeometric functions as
\begin{equation}\label{phi'larger}
\langle \phi'({\bf x}) \phi'({\bf y}) \rangle = \frac{\Lambda^3 \sigma^{-1}}{12 \pi}  \ _{1}F_{2} \left( \frac{3}{2};1,\frac{5}{2};-\frac{\Lambda^2 r^2 }{4} \right).
\end{equation} 

At close separations, we construct a smeared operator $\phi_{s}'$. The expectation value of this operator at zero separation is
\begin{equation}\label{phi'smallr}
\langle \phi_{s}'^{2} \rangle = \frac{\log(64) + 2\gamma -4}{2 \pi^2},
\end{equation}
where $\gamma=.577...$ is the Euler-Mascheroni constant. Smoothly
connecting the small $r$ (Eq.~\ref{phi'smallr}) and large $r$
(Eq.~\ref{phi'larger}) behavior as in Eq.~\ref{phiconnect}, the
$\Phi'$ (defined similarly to $z'$) correlator on the bubble at the time
of nucleation is:
\begin{eqnarray}\label{phi'2pt}
\langle  \Phi' \tilde{\Phi}' \rangle &=& \frac{(-Q)}{12 \pi} \left[ _{1}F_{2} \left( \frac{3}{2};1,\frac{5}{2};-\frac{R^2 \Psi^2}{4} \right) \right. \\ \nonumber & & \left. + \left(\frac{12 \gamma -6 \log(64) -24}{\pi} -1 \right)e^{-(R^2/2) \Psi^2} \right]
\end{eqnarray}

The integrals Eq.~\ref{Cl} and Eq.~\ref{Al} for the correlators
Eq.~\ref{eq-2pt} and Eq.~\ref{phi'2pt} must be evaluated
numerically. Calculation of the coefficients for every $l$ and $R$ is
unfortunately unfeasible because of the highly oscillatory behavior of
the integrands and sheer number of mode functions that must be
considered.  However, we have been able to deduce sufficiently good
approximate fits for $C_l$ and $A_l$ as a function of both $l$ and
$R$. In both cases, the power is dominated by a peak at $l\simeq R$.

The $C_l$ are nicely fit by the function
\begin{eqnarray}
 C_l &=& \frac{(-Q)}{(24 \pi)^2} \frac{83}{100} \frac{\sqrt{2}l + 1
 }{\sqrt{2} R +1} \\ \nonumber && \exp \left\{\frac{-1}{4R^2-2} \left[
   (\sqrt{2} l + 1 )^2 -(\sqrt{2} R + 1)^2 \right] \right\}.
\end{eqnarray}
The proposed fit for the $A_l$ consists of two power laws matched at
the $R=l$ peak. For $l \leq R$, the best fit is $A_l=(-Q) l^{5/4}/( 10
R^{9/4})$ and for $l>R$, the fit is $A_l=3 (-Q) R^{1.8} / (200 l^{2.6}
)$. Because these power law indices are slightly uncertain, we
only count the modes with $l \leq R$ in the $A_l$. This is
conservative, and also justified because these modes will not
contribute significantly to the sum in Eq.~\ref{Phiavg}.

With these initial conditions, we can now evolve each mode function
using Eq.~\ref{levolve} and then re-sum in Eq.~\ref{Phiavg} to find
the average size of the fluctuations at the turning point. We have
calculated $f$ and $g$ up to the $l$ corresponding to the last
unstable mode of the smallest $|Q|$, for all three inflation scales
($M_I=100$ GeV, $M_I=10^{14}$ GeV, and $M_I=10^{17}$ GeV). The results
for weak- and GUT-scale inflation are shown in Fig.~\ref{fg}, where
the dotted line indicates the boundary of the region over which the
perturbations become non-linear (non-linear to the left of the
line). It can be seen that in this model, even just quantum
perturbations of the bubble wall grow nonlinear in bubbles that start
at radii less than about one-tenth of the turnaround radius; this
grossly violates the assumption of spherical symmetry used in
tunneling calculations. On the other hand, none of the parameter space
in the high inflation scale case went non-linear, and at all scales
there is {\em always} a region of initial bubble radii near the
turnaround radius, for which nonlinearity never occurs.

\subsection{Thick Walls and Radiation}\label{TWRAD}
Just as we have no reason to expect a fluctuated region to be
spherically symmetric, we have no reason to assume that it will have
thin walls. An analysis of thick-walled true vacuum bubbles was
undertaken in Ref.~\cite{GVQ91,VV91}, where it was found that the
instabilities found in thin-walled case are still present in the form
of deformations normal to the bubble profile. In the case of small
false vacuum bubbles, there is no obvious consideration (such as a
corresponding instanton) to supply the profile of the bubble wall, and
so we can merely conjecture by precedent that the instability would be
retained in the thick-walled case as well.

Another consideration, applying to bubbles smaller than the false
vacuum horizon size, is whether inflation is spoiled by non-vacuum
contributions to the energy density. The perturbations on the bubble
wall translate into gravitational waves~\cite{II97,II99}, and since
the bound bubble solutions remain relatively close to their
gravitational radius and become distorted over many different length
scales on a relatively short time scale (see the quasi-exponential
growth in Fig.~\ref{usj}), they will be emitters of copious
gravitational radiation. Another problem arises if the kinetic and
gradient energy of the field becomes appreciable in the bubble
interior, either from intrusion of the wall (for example, imagine a
bubble being pinched in half by some non-linear perturbation), or from
particle production or other scalar modes propagating in from the
wall. If the emission of energy into the interior of the bubble from
any combination of these modes makes a significant contribution to the
equation of state, then inflation will not occur.

\section{Summary \& Discussion} \label{conclusions}

In this paper we have examined the feasibility of producing
inflationary universes from a spacetime with a small cosmological
constant. Reviewing the solutions allowed by the junction conditions,
there are several distinct possibilities if one allows for violations
of the null energy condition. The first is the Farhi, Guth \& Guven
(FGG) mechanism, in which (referring to Fig.~\ref{confdiags1} and Fig.~\ref{confdiags2}) the IV-I-II/III (solution 1) or IV-III-II/III (solution 2) solution is fluctuated
in the expanding phase and then tunnels to an unbound solution, as
shown in Fig.~\ref{tunnelst}. To the outside observer in region I of
the SdS conformal diagram it appears that the bubble has disappeared
behind the horizon, but on the other side of the wormhole, both an
inflationary universe and a non-inflating universe (an asymptotically
true-vacuum de Sitter region) have come into existence.  In one case,
shown in solution 5 of Fig.~\ref{confdiags2}, we can imagine an observer in
region III, then take the Schwarzschild mass $m\rightarrow 0$ limit
(thus removing regions I, II and IV entirely) and interpret the spacetime as that of the analytically continued Coleman-De Luccia instanton. Since the very same spacetime takes part in both the FGG mechanism and the Coleman-De Luccia false vacuum instanton, there may be some way to smoothly interpolate between these two processes;
we intend to explore this idea further in future work.

In addition to the FGG and regular instanton mechanisms, there are two
more possibilities. First, an inflationary universe might be directly
produced by some null energy condition (NEC) violating fluctuation
into one of the unbound or monotonic solutions shown in Solutions 5-13 of
Fig.~\ref{confdiags2}. To the outside observer in region I of the SdS
conformal diagram, these solutions would be indistinguishable from the
fluctuation of a black hole, but they would secretly entail the
creation of everything on the other side of the wormhole, as in the
FGG tunneling. A second method of direct production is the fluctuation
(which does not require an NEC violation) of the IV-I-II'/III-II
solution (solution 3 and 4). These solutions only exist in the case where the
interior and exterior cosmological constants are comparable, and so
would not correspond to the nucleation of inflation from a universe
like ours, but they might be of interest in understanding transitions
between nearly degenerate vacua.

Examining these classical solutions to first order, we have shown that
an instability to aspherical perturbations exists in those solutions
which possess a turning point. This includes the bound (solutions 1 and 2)
and unbound solutions (solutions 5-9). In the latter case there is
no clear way to set an initial radius or initial perturbation
amplitude, so we can say only that collapsing bubbles are violently
unstable. The bound solutions are amenable to quantitative
investigation, and we have focused on the growth of perturbations in
the expanding phase that precedes tunneling in the FGG mechanism. For
bound expanding bubbles formed by the fluctuations of a scalar field
in de Sitter space, there is no instanton to enforce spherical
symmetry, so we would expect initial aspherical perturbations to be
relatively large. Since there is no detailed model for the fluctuating
scalar field to see {\em how} large, we have instead calculated an
estimate of the minimal deviations from spherical symmetry in light of
quantum fluctuations, and present this as an extremely rare, best case
scenario for spherical symmetry. These minimal fluctuations were then
evolved to the turning point of the bound solutions, which is the
point in the FGG mechanism where there is a chance for tunneling to an
inflationary universe to occur. Of the three representative energy
scales for inflation (false vacuum energy densities) we have studied,
the evolved minimal perturbations on a Plank scale bubble remain small
over most of the allowed parameter space, while the perturbations on
GUT and weak scale bubbles can grow nonlinear if they start at a
sufficiently small ($\alt 10\%$) fraction of the turnaround radius.
Thus even in the best-case scenario some bubbles become nonlinear, but
on the other hand there will always in principle be some that do not.

The instability introduces complications into the use of the FGG
mechanism as a means of baby universe production. The existing
calculations of the tunneling rate rely heavily on the assumption of
spherical symmetry. It is unclear how to perform a similar calculation
for a (possible non-linearly) perturbed bubble, as the number of
degrees of freedom has drastically increased and the assumption of a
minisuperspace of spherically symmetric metrics is no longer
good. Further, the bubble interior will become filled with scalar
gradient and kinetic energy and gravity waves, possibly upsetting the
interior sufficiently to prevent inflation. One might argue that in an
eternal universe there is plenty of time to wait around for a
fluctuation which is sufficiently spherical. However, to fully
understand the importance of the FGG mechanism, one must both have a
model of the scalar field fluctuations, which would predict the
distribution of bubble shapes and masses, and also a model for
tunneling in the presence of asphericities. In addition, one must
understand the competition among the various processes which result in
the formation of inflating false vacuum regions, and thus show that
the FGG mechanism is not a relative rarity.  These problems represent
significant calculations in a theory of quantum gravity which we do
not yet possess, but progress in these areas would undoubtedly improve
our understanding of the initial conditions for inflation.

\begin{acknowledgments}
The authors wish to thank A. Albrecht, T. Banks, P.J. Fox, S. Gratton, and T. Mai for their assistance in the development of this work.
\end{acknowledgments}

\suppressfloats

\end{document}